\documentclass{jfm}

\usepackage{graphicx}
\usepackage{natbib}
\usepackage{epstopdf}
\usepackage{amsmath}
\usepackage{rotating}

\usepackage{ amssymb, amsfonts}

\usepackage{xcolor}

\usepackage{subfig}

\ifCUPmtlplainloaded \else
  \checkfont{eurm10}
  \iffontfound
    \IfFileExists{upmath.sty}
      {\typeout{^^JFound AMS Euler Roman fonts on the system,
                   using the 'upmath' package.^^J}%
       \usepackage{upmath}}
      {\typeout{^^JFound AMS Euler Roman fonts on the system, but you
                   dont seem to have the}%
       \typeout{'upmath' package installed. JFM.cls can take advantage
                 of these fonts,^^Jif you use 'upmath' package.^^J}%
      }
  \else
  \fi
\fi

\ifCUPmtlplainloaded \else
  \checkfont{msam10}
  \iffontfound
    \IfFileExists{amssymb.sty}
      {\typeout{^^JFound AMS Symbol fonts on the system, using the
                'amssymb' package.^^J}%
       \usepackage{amssymb}%
       \let\le=\leqslant  \let\leq=\leqslant
       \let\ge=\geqslant  \let\geq=\geqslant
      }{}
  \fi
\fi

\ifCUPmtlplainloaded \else
  \IfFileExists{amsbsy.sty}
    {\typeout{^^JFound the 'amsbsy' package on the system, using it.^^J}%
     \usepackage{amsbsy}}
    {\providecommand\boldsymbol[1]{\mbox{\boldmath $##1$}}}
\fi

\newsavebox{\astrutbox}
\sbox{\astrutbox}{\rule[-5pt]{0pt}{20pt}}

\newcommand{\dd}{\ensuremath{\mathrm{d}}}
\newcommand{\id}{\ensuremath{\hphantom{.}\mathrm{d}}}
\newcommand{\f}[2]{\ensuremath{\frac{#1}{#2}}} 
\newcommand{\df}[2]{\ensuremath{\f{\dd #1}{\dd #2}}}

 \mathchardef\mhyphen="2D

\definecolor{myred}{RGB}{217, 10, 10}
\definecolor{myblue}{RGB}{0, 0, 230}
\definecolor{LGrey}{rgb}{.5,.5,.5}

\title[Varying roughness density in turbulent channel flow]{
Turbulent flow over transitionally rough surfaces with varying roughness densities}

\author[M. MacDonald, L. Chan, D. Chung, N. Hutchins and A. Ooi]
{M. MacDonald\thanks{Email address for correspondence: michael.macdonald@unimelb.edu.au},\ns
L. Chan,
D. Chung,
N. Hutchins
and A. Ooi}

\affiliation{Department of Mechanical Engineering, University of Melbourne, Victoria 3010, Australia}

\pubyear{2015}
\volume{999}
\pagerange{998--999}
\date{?; revised ?; accepted ?}
\begin{document}

\maketitle

\begin{abstract}
We investigate rough-wall turbulent flows through direct numerical simulations of flow over three-dimensional transitionally rough sinusoidal surfaces. The roughness Reynolds number is fixed at $k^+=10$, where $k$ is the sinusoidal semi-amplitude, and the sinusoidal wavelength is varied, resulting in the roughness solidity, $\Lambda$ (frontal area divided by plan area) ranging from 0.05 to 0.54. The high cost of resolving both the flow around the dense roughness elements and the bulk flow is circumvented by the use of the minimal-span channel technique, recently demonstrated by Chung \emph{et al.} (\emph{J.\ Fluid Mech.}, vol.\ 773, 2015, pp.\ 418--431) to accurately determine the Hama roughness function, $\Delta U^+$. Good agreement of the second-order statistics in the near-wall roughness-affected region between minimal- and full-span rough-wall channels is observed. 
In the sparse regime of roughness ($\Lambda \lesssim 0.15$) the roughness function increases with increasing solidity, while in the dense regime the roughness function decreases with increasing solidity. It was found that the dense regime begins when $\Lambda\gtrsim0.15$--0.18, in agreement with the literature. 
A model is proposed for the limit of $\Lambda\rightarrow\infty$, which is a smooth wall located at the crest of the roughness elements. This model assists with interpreting the asymptotic behaviour of the roughness, and the rough-wall data presented in this paper show that the near-wall flow is tending towards this modelled limit.
The peak streamwise turbulence intensity, which is associated with the turbulent near-wall cycle, is seen to move increasingly further away from the wall with increasing solidity. In the sparse regime, increasing $\Lambda$ reduces streamwise turbulent energy associated with the near-wall cycle, while increasing $\Lambda$ in the dense regime increases turbulent energy.
An analysis of the difference of the integrated mean-momentum balance between smooth- and rough-wall flows reveals that the roughness function decreases in the dense regime due to a reduction in the Reynolds shear stress. This is predominantly due to the near-wall cycle being pushed away from the roughness elements, which leads to a reduction in turbulent energy in the region previously occupied by these events.

\end{abstract}


\section{Introduction}

Turbulent flow over a rough wall is of critical interest in many engineering and geophysical wall-bounded flows, as the wall cannot generally be considered to be smooth. The inclusion of roughness alters the flow dynamics and increases drag, which is most readily quantified in terms of the (Hama) roughness function, $\Delta U^+$. The relevant question in rough-wall flows is therefore how the roughness function is related to the surface geometry, which has been an area of intense research with a multitude of surfaces being tested (refer to reviews by \citealt{Jimenez04} and \citealt{Flack10}). The rough surface alters the near-wall behaviour of the turbulent flow, resulting in the log- and outer-layers of the mean streamwise velocity being shifted down by $\Delta U^+$, when compared to smooth-wall flow at matched friction Reynolds number $Re_\tau=U_\tau h /\nu$; here $U_\tau=\sqrt{\tau_w/\rho}$ is the friction velocity, $\tau_w$ is the wall shear stress, $\rho$ is density, $h$ is the channel half-height or pipe radius, and $\nu$ is the kinematic viscosity. Starting from the log-law for smooth- and rough-wall flows at matched friction Reynolds numbers, the roughness function can be directly related to the difference of the skin friction coefficients \citep{Schultz09},
\begin{equation}
\label{eqn:DU_CF}
\Delta U^+(Re_\tau) = \sqrt{\frac{2}{C_{fs}(Re_\tau)}}-\sqrt{\frac{2}{C_{fr}(Re_\tau)}},
\end{equation}
where subscripts $s$ and $r$ are used to denote the smooth- and rough-wall flows, respectively. Here, the skin friction coefficient $C_f=2\tau_w/U_h^2=2/{U_h^+}^2$ is defined on the streamwise centreline velocity $U_h=U(z=h)$  (e.g. \citealt{Mayoral11jfm}). It is important to emphasise that this relationship only holds when the friction Reynolds number is kept constant in both the smooth- and rough-wall flows.

\setlength{\unitlength}{1cm}
\begin{figure}
\centering
 \captionsetup[subfigure]{labelformat=empty}

	\subfloat[]{
		\includegraphics[width=0.49\textwidth,trim=0 0 0 0,clip=true]{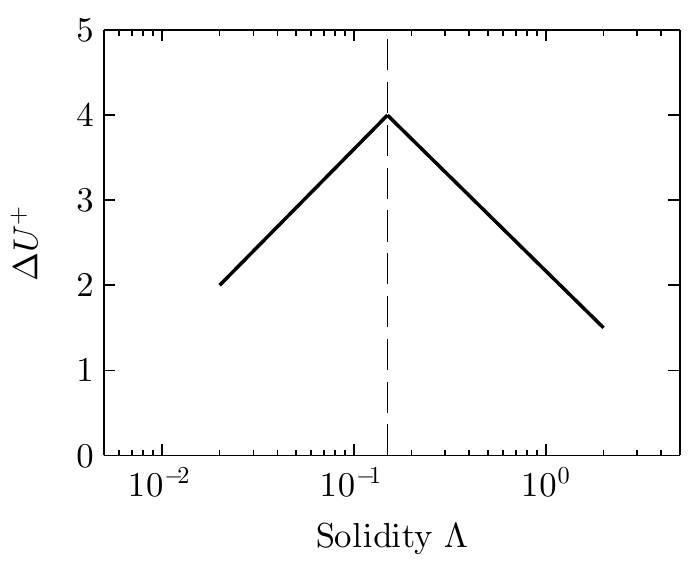}
		\label{fig:UxValid1}
	}
	\subfloat[]{
		\includegraphics[trim=0 -40 0 0,clip=true,width=0.49\textwidth]{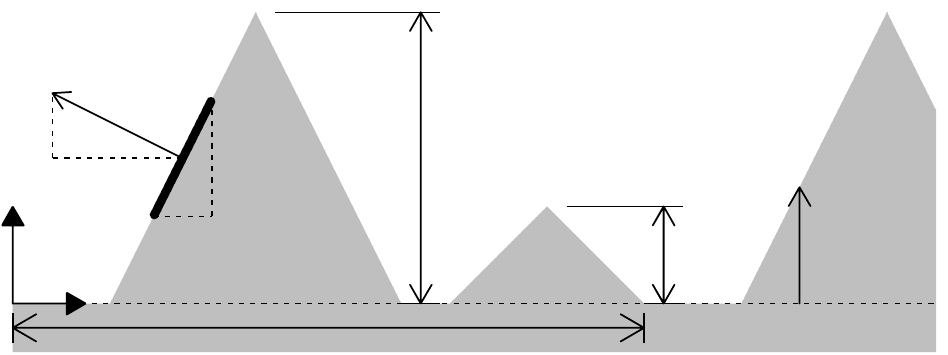}
		\label{fig:UxValid2}
	}
	\put(-13.6,5.0){(\emph{a})}
	\put(-12,4.6){Sparse}
	\put(-12.25,4.0){\includegraphics{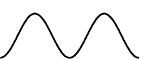}}
	\put(-9,4.6){Dense}
	\put(-9.25,4.0){\includegraphics{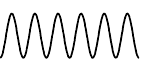}}
	\put(-6.9,5.0){(\emph{b})}
	\put(-3.5,4.8){Flow}
	\put(-1.05,1.8){$z_w(x)$}
	\put(-3.65,2.25){$k_1$}
	\put(-1.95,1.55){$k_2$}
	\put(-6.15,2.75){$\hat{\mathbf{n}}$}
	\put(-6.73,2.45){$\hat{n}_z$}
	\put(-6.2,2.0){$\hat{n}_x$}
	\put(-5.7,2.75){$\dd A$}
	\put(-5.2,2.2){$\dd A_f$}
	\put(-5.65,1.65){$\dd A_p$}
	\put(-4.5,0.8){$\lambda$}
	\put(-6.85,1.95){$z$}
	\put(-6.15,1.4){$x$}
	\put(-3.5,3.9){$ES = \dfrac{2k_1+2k_2}{\lambda}$}
	\put(-3.2,3.1){$\Lambda = \dfrac{k_1+k_2}{\lambda}$}
	\put(-4.0,4.6){\vector(1,0){1.5}}
	\vspace{-2.2\baselineskip}
\caption{(\emph{a}) Sketch of roughness function $\Delta U^+$ against solidity $\Lambda$, based on figure 1(\emph{a}) of \cite{Jimenez04}. Vertical dashed line at $\Lambda=0.15$ demarcates the sparse and dense regimes. (\emph{b}) Sketch of a two-dimensional rough surface in streamwise--wall-normal plane, with repeating unit wavelength $\lambda$ and roughness element heights $k_1$ and $k_2$.
}
	\label{fig:DUSketch}
\end{figure}
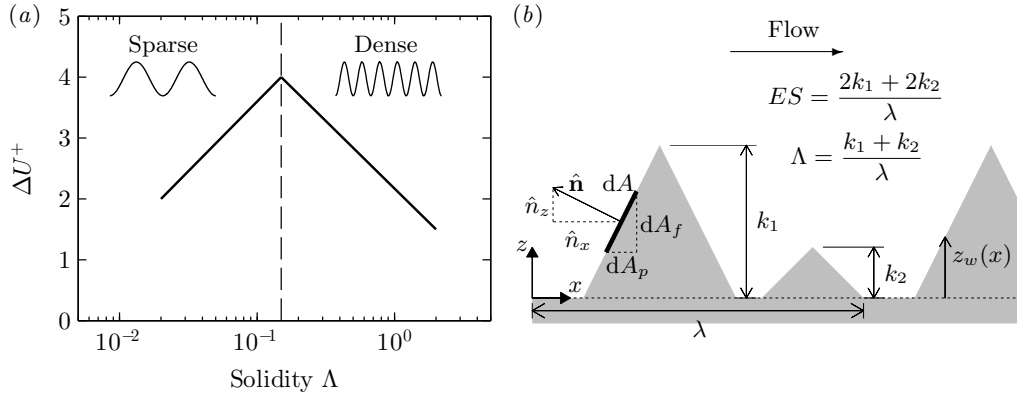

It has become clear that the roughness function depends on the density of the roughness. The roughness density is often described by the frontal solidity, $\Lambda$, first defined by \cite{Schlichting36} as $S_f/S_p$, where $S_f$ is the total frontal-projected area of the roughness in the direction perpendicular to the flow and $S_p$ is the total plan area of the roughness in the direction parallel to the flow. 
There is some ambiguity as to how the frontal area $S_f$ is defined, particularly for complex, non-uniform rough surfaces. Some authors (for example, \citealt{Placidi15}) only include the frontal area of roughness elements that are not sheltered by other elements upstream of it in a repeating unit. Alternatively, the definition proposed by \cite{Schlichting36} includes the total frontal area of every element, and no distinction is made between sheltered and unsheltered elements. To distinguish these two definitions, figure \ref{fig:DUSketch}(\emph{b}) gives an example of a two-dimensional rough surface composed of triangular bars aligned in the spanwise direction. The repeating unit has length $\lambda$ and is comprised of two triangles, one of height $k_1$ and a second downstream of the first with height $k_2<k_1$. The first definition of $S_f$ would only include the first roughness element of height $k_1$, as the second element in the repeating unit is sheltered by the first and hence is not exposed to the flow. The second definition includes all elements in a repeating unit, so that $S_f = k_1+k_2$. It is not the purpose of this paper to compare different measures of roughness density so henceforth the second definition of $S_f$ will be used. To rigorously define the solidity, consider an elemental cell of the rough wall with area $\dd A$ and local unit wall-normal vector $\hat{\mathbf{n}}=(\hat{n}_x,\hat{n}_y,\hat{n}_z)$, for the streamwise, spanwise, and wall-normal directions (figure \ref{fig:DUSketch}\emph{b}). The plan elemental area is then $\dd A_p = \hat{n}_z \dd A$ and the frontal area is $\dd A_f = \hat{n}_x \dd A$. The total plan area of the surface is therefore defined as $S_p = \int \hat{n}_z \dd A$ and the total frontal area is $S_f = \int_{\hat{n}_x<0} -\hat{n}_x \dd A = \tfrac{1}{2}\int |\hat{n}_x |\dd A$. Hence the solidity can be precisely defined,
\begin{equation}
\label{eqn:sol}
\Lambda = \frac{S_f}{S_p} = \frac{1}{2}\frac{\int|\hat{n}_x|\dd A}{\int\hat{n}_z\dd A}.
\end{equation}
A solidity of $\Lambda = 0$ represents a smooth wall, while the limit of $\Lambda\rightarrow\infty$ for a roughness in which the ratio of solid volume to fluid volume remains constant (e.g. sinusoidal roughness) represents dense needle-like roughness.
\cite{Napoli08} introduced the concept of effective slope, $ES$, as an alternative to the roughness solidity, which is calculated based on the gradient of the wall-location $z_w$,
\begin{equation}
ES = \frac{1}{L_xL_y}\int_{0}^{L_y}\int_{0}^{L_x} \left| \frac{\partial z_w(x,y)}{\partial x}\right|\id x\id y = 2\Lambda,
\end{equation}
where $L_x$ and $L_y$ are the streamwise and spanwise lengths of the rough surface, respectively. Noting that $\hat{n}_z \dd A = \dd x\dd y$, then the roughness plan area is $S_p=\int\hat{n}_z\dd A=L_x L_y$, while the frontal area is $S_f = \tfrac{1}{2}\int|\hat{n}_x|\dd A = \tfrac{1}{2}\int\int|\hat{n}_x|/\hat{n}_z\dd x\dd y = \tfrac{1}{2}\int \int |n_x|/ n_z \dd x\dd y$. 
The vector normal to the wall is $\mathbf{n} =(-\partial z_w/\partial x,-\partial z_w/\partial y,1)$,
and hence it can be seen that the solidity (\ref{eqn:sol}) is half the effective slope, $\Lambda \equiv \tfrac{1}{2}ES$. More complex definitions of roughness density have been proposed  which use measures of the roughness elements arrangement and orientation (e.g. \citealt{Waigh98}), however in this study the solidity as defined in (\ref{eqn:sol}) will be used.

It has been observed (figure \ref{fig:DUSketch}\emph{a}) that the roughness function increases with density for $\Lambda<0.15$ \citep{Jimenez04,Flack14}. Above this point, the roughness function then decreases with increasing density. It has often been qualitatively described that increasing the roughness density while in the sparse regime ($\Lambda < 0.15$) increases drag due to the increased frontal area of the roughness. In the dense regime ($\Lambda > 0.15$), mutual sheltering of roughness elements leads to a decrease in drag as the density is increased \citep{Oke88,Macdonald98,Jimenez04}. However, this description does not appear to have been studied in a quantitative manner. 
In meteorology, these regimes have been investigated in more depth. The close proximity of buildings in urban canopies often results in the solidity taking values of $0.1$--$0.7$
 \citep{Grimmond99}, a range covering both the sparse and dense regimes. 
 Here, the terminology is slightly different, with the sparse and dense regimes being referred to as the wake and skimming regimes, respectively. Conceptual models of the skimming, or dense, regime portray stable vortices below the crest of the roughness with high-speed fluid flowing over the top; this is not dissimilar to those conceptual models describing the so-called $d$-type roughness \citep{Jimenez04}.
Various efforts have been made to predict the resultant drag on the rough wall based on the solidity (e.g., \citealt{Raupach94,Macdonald98,Lien05}) or on more complex measures of the roughness density (e.g. \citealt{Waigh98,Flack10}). 
\cite{Grimmond99} reviewed several different models in an effort to compare their performance, however the limited accuracy of field and experimental data at the time made comparisons difficult. Estimating the friction velocity $U_\tau$ and hence $\Delta U^+$ in experimental studies over rough walls is challenging, leading to errors of up to $\pm10\%$ \citep{Schultz09}, making analyses and predictions difficult.
More recently, \cite{MillwardHopkins11} and \cite{Yang16} have formulated drag-prediction models that use a sheltering argument to account for the interaction between roughness elements.
While these models show promise for predicting drag, the purpose of this paper is to investigate and understand the physics of the sparse and dense regimes which could hopefully motivate future models.

\cite{Napoli08} conducted Direct Numerical Simulations (DNSs) of two-dimensional sinusoidal roughness with multiple modes. They showed that the roughness function increases with effective slope until $ES\approx 0.4$, after which point it appears independent of $ES$. \cite{Schultz09} conducted experiments of flow over pyramids in which the amplitude and slope was systematically varied. They showed, in light of their data and the results of \cite{Napoli08}, that the roughness function scales on $ES$ for $ES\lesssim0.35$, which they termed the waviness regime. Above this point, there was little dependence on the effective slope, which they termed the `roughness' regime, as the data more readily scaled on the roughness Reynolds number $k^+=k h/\nu$; here $k$ is the roughness height. The demarcation point of $ES\approx0.35\Rightarrow\Lambda\approx0.175$ is similar to the transition to dense roughness, $\Lambda\approx0.15$. However, \cite{Napoli08} only observed a slight reduction in the roughness function for the two dense roughness cases that they simulated, with solidities of $\Lambda=0.275$ and 0.38. This may be because these solidity values are not sufficiently large enough to convincingly see a reduction, as well as these cases enter the fully rough regime ($\Delta U^+\approx10$). The sparse regime cases that were simulated are more likely thought of as being in the transitionally rough regime ($\Delta U^+\lesssim8$), making it difficult to isolate just the effects of roughness density.
 Moreover, their rough surface  was created by the superposition of multiple sinusoidal modes. The dense cases had very large roughness peaks that were relatively sparse, however these peaks would have had a dominant effect on the roughness function. This idea is supported by the work of \cite{Hagishima09}, who conducted an experimental study of cubes in different arrangements and with varying heights. In particular, the authors analysed roughness arrays which had cubes of two different heights and varied the arrangement of the cubes with the shorter heights. They showed that varying the arrangement of the shorter roughness elements has little effect on the total drag. In this case, perhaps a more useful measure is the solidity according to \cite{Macdonald00}, which is weighted towards the taller roughness elements.
This may explain why \cite{Napoli08} did not see a reduction in $\Delta U^+$, as the density of the peaks alone is more important than the behaviour of the roughness in the troughs.  To simplify this current study a uniform roughness height is used in order to systematically analyse the dense regime of roughness.

Most other studies in the dense roughness regime involve various arrangement of cubes with a regular height. \cite{Leonardi10} used DNS to investigate turbulent flow over cube roughness, with the solidity ranging from 0.04 to 0.25. The authors saw a reduction in $\Delta U^+$ for solidity values greater than $0.15$. By analysing the total friction and pressure drag forces  on the rough wall, it was found that the dense roughness regime had a frictional drag that was about 5\% of the total drag. 
\cite{Leonardi07} investigated $d$- and $k$-type roughness by simulating flow over two-dimensional longitudinal bars in which the spacing between the bars was varied, with the solidity ranging from 0.02 to 1. Even though the authors were not focussed on the sparse and dense roughness regimes, the reduction in the roughness function is seen for solidity values greater than $0.11$--$ 0.33$. However, the bulk velocity was kept constant across these low Reynolds number simulations so that the friction velocity varies, which makes isolating the effects of friction and roughness Reynolds numbers on $\Delta U^+$ difficult.

In the current study, three-dimensional sinusoidal roughness is used and the roughness Reynolds number remains fixed at $k^+=10$ while the wavelength
and hence solidity varies.
The equivalent sand grain roughness $k_s$ is a factor that relates a length scale of a rough surface to the sand grain data of \cite{Nikuradse33} to
ensure a collapse of the roughness function in the fully rough regime. For a particular sinusoidal surface with $\Lambda=0.18$, it was found to be $k_s/k=4.1$ in \cite{Chan15},
meaning $k_s^+\approx41$ for this surface.
 The equivalent sand grain roughness is not known for for the other sinusoidal surfaces tested in this study ($0.05\le\Lambda\le0.54$).
It will likely be unique for each case but of similar magnitude to the $\Lambda=0.18$ case,  which places them in the transitionally rough regime \citep{Flack10}. Many industrial
systems operate with roughness in the transitional regime, such as on ship hulls and turbine blades \citep{Flack12}, making understanding this
regime of importance to engineers.
Fully rough flows are not the focus of this paper, although where possible, comparisons to fully rough flows are made.

Cube roughness studies \citep{Kanda04,Coceal06,Leonardi10} often have fixed cube dimensions, and these cubes are moved closer together to increase solidity. The ratio of solid (cube) volume to fluid volume therefore changes with solidity. A solidity of unity implies that all the cubes touching  one another, so that there is a  smooth wall located at the top surface of the cubes. In the present study, even in the limit of $\Lambda\rightarrow\infty$, the ratio of solid to fluid volume remains unchanged. 
Conceptually, once this sinusoidal roughness is dense enough, the roughness would more readily be thought of as a bed of tightly packed `pins' that protrude into the fluid.
Moreover, in studies and models of cube roughness, an additional solidity is defined based on the plan area of the top surface of the cube and the plan area of the repeating element. This plan solidity concept is only useful for surfaces where there are two distinct roughness heights, such as the underlying wall and the top surface of the cube. In more complicated rough walls, such as sinusoidal roughness, the plan area is not well defined and hence this plan solidity is not reported here.

DNS of the dense regime of roughness is challenging because both the dense roughness elements and the bulk flow must be resolved. If we consider a full-span rough-wall channel with 24 cells defining a roughness element in the streamwise and spanwise directions, then the grid for the densest roughness case in this study would have approximately 340 million cells in total. This represents a large investment of computational resources for a single roughness case which could be better deployed to simulating several different roughness densities. This could be achieved using the minimal-span channel framework, which circumvents the prohibitive cost of resolving the bulk flow. The minimal channel was first studied by \cite{Jimenez91}, with further work being conducted by various authors \citep{Jimenez99,Hamilton95,Flores10, Hwang13,Lozano14}. Importantly, the minimal channel concept was seen to preserve the near-wall behaviour of the turbulent flow, for both the buffer layer \citep{Jimenez91} and the log layer \citep{Flores10}. Because roughness is thought to predominantly alter only the near-wall flow, then it follows that the minimal channel approach can be used to study rough-wall behaviour. Recently, \cite{Chung15} demonstrated that using a channel with a spanwise length of a few hundred wall-units allowed the roughness function $\Delta U^+$ to be estimated at a reduced cost, when compared to full-span channels which are conventionally used. As a comparison, for the previously mentioned densest roughness case, the full-span channel would require 340 million cells as compared to a requirement of only 21 million cells in a minimal-span channel, representing a full order of magnitude saving in the grid.

The streamwise, spanwise, and wall-normal directions are denoted by $x$, $y$, and $z$, respectively with corresponding velocity components $u$, $v$, and $w$. Quantities with an overbar, e.g., $\overline{u}$, are time averaged, while angle brackets, $\langle u \rangle$, denote the spatial average in a plane of constant $z$, defined as
\begin{equation}
\langle u\rangle(z_i) = \frac{\sum_{i\in \text{fluid}} u_i \mathcal{V}_i}{\Delta z L_x L_y},
\end{equation}
where $\mathcal{V}_i$ is the volume of fluid in cell $i$ which falls between $z-\Delta z/2 < z_i < z+\Delta z/2$ and $L_x$ and $L_y$ are the streamwise and spanwise domain lengths. This is the superficial spatial average \citep{Breugem06}, as the denominator is the total volume of both fluid and solid parts.
 Upper case variables indicate both spatial and temporal averaging, i.e., $U = \langle \overline{u} \rangle$. 
 The superscript $+$ is used to denote variables non-dimensionalised on viscosity, $\nu$, and friction velocity $U_\tau=\sqrt{\tau_w/\rho}$, where $\tau_w$ is the temporally and spatially averaged wall drag per plan area and $\rho$ is the fluid density. Velocity fluctuations, $u'$, are defined based on the difference between the instantaneous velocity and the temporally and spatially averaged velocity at a given wall-normal location, $u'(x,y,z,t) \equiv\langle\overline{u}\rangle(z) -u(x,y,z,t)$. 
 Root-mean-square (RMS) velocity fluctuations are defined as $u'_{rms}(z) =\langle\overline{u'^2}\rangle^{1/2}$.

In section \ref{sect:numProc}, the numerical procedure will be briefly outlined. Following that, validation of the minimal-span channel is performed by comparing first- and second-order statistics with full-span channel data (section \ref{sect:minspanvalid}). A model for the limit of $\Lambda\rightarrow\infty$ is then described in section \ref{sect:dense}, which will be used to assist in the interpretation of data in the dense regime of roughness.
In section \ref{sect:rough}, the rough-wall results will be analysed in detail, with a quantitative explanation given for the decrease in the roughness function that is observed in the dense regime. Conclusions are  offered in section \ref{sect:discConc}.


\section{Numerical procedure}
\label{sect:numProc}
The numerical procedure used herein has been described and validated in \cite{Chan15}, but the important features are repeated here. The Navier--Stokes equations for incompressible flow are solved in Cartesian coordinates,
\begin{equation}
\label{eqn:navierStokes}
	\nabla\cdot{\mathbf{u}} = 0,
\quad \quad
\frac{\partial \mathbf{u}}{\partial t} + \mathbf{u}\cdot\nabla\mathbf{u} = -\f{1}{\rho}\nabla p+\nu\nabla^2\mathbf{u}+\mathbf{G},
\end{equation}
where $\mathbf{u}=(u,v,w)$, 
$t$ is time,
$p$ is pressure,
and 
$\mathbf{G}=(G_x(t),0,0)$ is the spatially uniform, time-dependent pressure-gradient that drives the flow at constant mass flux.
These equations are solved using CDP, a second-order finite-volume code \citep{Ham04,Mahesh04}.
Periodic boundary conditions are applied in the streamwise and spanwise directions, while the no-slip and impermeability conditions are applied on the channel walls at $z=0+z_w$ and $z=2h+z_w$.  
The rough surface is defined by a three-dimensional sinusoidal function,
\begin{equation}
\label{eqn:roughWall}
z_w = k\cos\left(\frac{2\pi x}{\lambda_x}\right)\cos\left(\frac{2\pi y}{\lambda_y}\right),
\end{equation}
where $k$ is the mean-to-peak (semi) amplitude of the roughness, and $\lambda_x$ and $\lambda_y$ are the streamwise and spanwise wavelengths. In this study, these are kept the same, so that $\lambda = \lambda_x=\lambda_y$.
All simulations have a constant amplitude of $k^+ = 10$, corresponding to a blockage ratio of $k/h=1/18$ at $Re_\tau=180$. 
The wavelength $\lambda$ is varied, resulting in the roughness density and hence steepness of the surface changing.
The solidity, shown to be exactly half the effective slope, therefore increases as $\lambda$ decreases.
For the current sinusoidal surface, the solidity can be evaluated as $\Lambda = (4/\pi) k/\lambda_x$ \citep{Chan15}.

\begin{table}
\centering
\resizebox{\textwidth}{!}{
\begin{tabular}{ c |c c c | c c c  | c c c c | c c c c | c c}
$Re_\tau$ 	& $k^+$	& $\lambda^+$  & $\Lambda$	& $L_x/h$ 	&  $L_x^+$		& $L_y^+$	& $N_x$	& $N_{\lambda}$	& \multicolumn{2}{c}{$N_y$$\times$$N_z$}& $\Delta x^+$	& $\Delta y^+$	& $\Delta z_w^+$	& $\Delta z_h^+$  & $U_b^+$	& $\Delta U^+$\\[-0.4em]
\hline\multicolumn{17}{c}{}\\[-1.2em]
\multicolumn{17}{c}{Full-span pipe, predominantly from \cite{Chan15}} \\[-0.4em]
\hline \\[-3.5ex]
180	& -	& -	& 0	& $4\pi$	& 2265	& 1133	& 384	& -	& \multicolumn{2}{c|}{13680}	& 5.9	& 4.7	& 0.4	& 4.8	& 15.0 & -	\\
180	& 10	& 283	& 0.05	& $4\pi$	& 2265	& 1132	& 512	& 64	& \multicolumn{2}{c|}{19872}	& 4.4	& 3.9	& 0.4	& 4.0& 12.6 & 2.64	\\
180	& 10	& 188	& 0.07	& $4\pi$	& 2260	& 1130	& 512	& 43	& \multicolumn{2}{c|}{19872}	& 4.4	& 3.9	& 0.4	& 4.0& 11.8 & 3.40	\\
180	& 10	& 141	& 0.09	& $4\pi$	& 2258	& 1129	& 512	& 32	& \multicolumn{2}{c|}{19872}	& 4.4	& 3.9	& 0.4	& 4.0& 11.5 & 3.69	\\
180	& 10	& 113	& 0.11	& $4\pi$	& 2264	& 1132	& 512	& 26	& \multicolumn{2}{c|}{19872}	& 4.4	& 3.9	& 0.4	& 4.0& 11.2 & 3.91	\\
180	& 10	& 94	& 0.14	& $4\pi$	& 2265	& 1133	& 640	& 27	& \multicolumn{2}{c|}{24768}	& 3.5	& 2.9	& 3.5	& 3.0 & 11.1 & 4.06 	\\
182	& 10	& 71	& 0.18	& $4\pi$	& 2291	& 1146	& 512	& 16	& \multicolumn{2}{c|}{24864}	& 4.5	& 1.7	& 0.4	& 3.5& 11.1 & 4.10	\\
179	& 10	& 47	& 0.27	& $4\pi$	& 2251	& 1126	& 1152	& 24	& \multicolumn{2}{c|}{102144}	& 2.0	& 1.7	& 0.2	& 1.7& 12.8 & 3.64	\\[-0.4em]
\hline
\hline
 $Re_\tau$ 	& $k^+$	& $\lambda^+$  & $\Lambda$	& $L_x/h$ 	&  $L_x^+$		& $L_y^+$	& $N_x$	& $N_{\lambda}$	& $N_y$	& $N_z$& $\Delta x^+$	& $\Delta y^+$	& $\Delta z_w^+$	& $\Delta z_h^+$  & $U_b^+$ & $\Delta U^+$\\[-0.4em]
\hline\multicolumn{17}{c}{}\\[-1.2em]
\multicolumn{17}{c}{Full-span channel} \\[-0.4em]
\hline\\[-3.5ex]
180	& -	& -	& 0	& $4\pi$	& 2260	& 1130	& 453	& -	& 226	& 156	& 5.0	& 5.0	& 0.3	& 5.1	& 16.1 &  -	\\
180	& 10	& 113	& 0.11	& $4\pi$	& 2257	& 1128	& 453	& 24	& 226	& 211	& 5.0	& 5.0	& 0.3	& 4.8	&12.4 & 3.72	\\
181	& 10	& 71	& 0.18	& $4\pi$	& 2275	& 1138	& 510	& 24	& 254	& 156	& 4.5	& 4.5	& 0.3	& 5.2	& 12.3 & 3.93	\\[-0.5em]
\hline\multicolumn{16}{c}{}\\[-1.2em]
\multicolumn{17}{c}{Minimal-span channel} \\[-0.4em]
\hline\\[-3.5ex]
180	& -	& -	& 0	& $4\pi$	& 2263	& 113	& 384	& -	& 32	& 192	& 5.9	& 3.5	& 0.2	& 4.0	& 17.4& -	\\
182	& -	& -	& 0	& $4\pi$	& 2281	& 143	& 512	& -	& 32	& 160	& 4.5	& 4.5	& 0.3	& 4.9	& 17.2 & -	\\
180	& -	& -	& 0	& $2\pi$	& 1133	& 142	& 256	& -	& 32	& 160	& 4.4	& 4.4	& 0.3	& 4.9	& 17.0	& -	\\
180	& 10	& 113	& 0.11	& $4\pi$	& 2258	& 113	& 480	& 24	& 24	& 128	& 4.7	& 4.7	& 0.4	& 6.1& 13.7	& 3.60	\\
182	& 10	& 71	& 0.18	& $2\pi$	& 1143	& 143	& 384	& 24	& 48	& 128	& 3.0	& 3.0	& 0.4	& 6.1	& 12.9& 4.14	\\
181	& 10	& 47	& 0.27	& $4\pi$	& 2274	& 142	& 1152	& 24	& 72	& 128	& 2.0	& 2.0	& 0.4	& 6.1	& 13.1& 3.97	\\
179	& 10	& 47	& 0.27	& $2\pi$	& 1122	& 140	& 576	& 24	& 72	& 128	& 1.9	& 1.9	& 0.4	& 6.0	& 13.0 &3.80	\\
180	& 10	& 35	& 0.36	& $2\pi$	& 1134	& 142	& 768	& 24	& 96	& 128	& 1.5	& 1.5	& 0.4	& 6.1	&13.1& 3.81	\\
179	& 10	& 24	& 0.54	& $2\pi$	& 1125	& 141	& 1152	& 24	& 144	& 128	& 1.0	& 1.0	& 0.4	& 6.0	&13.3& 3.44	\\
181	& 10	& 0	& $\infty$	& $4\pi$	& 2279	& 142	& 512	& -	& 32	& 160	& 4.5	& 4.5	& 0.3	& 4.6	& 16.5&0.93	\\[-0.5em]
\hline\\[-3.7ex]
359	& -	& -	& 0	& $2\pi$	& 2258	& 113	& 384	& -	& 32	& 224	& 5.9	& 3.5	& 0.4	& 7.2	& 21.9& -	\\
360	& 10	& 113	& 0.11	& $2\pi$	& 2260	& 113	& 480	& 24	& 24	& 256	& 4.7	& 4.7	& 0.3	& 6.6 &18.6	& 3.25	\\
\end{tabular}
}
\vspace{-0.3\baselineskip}
\caption{Description of the different simulations performed. 
$k$ is the mean-to-peak (semi) amplitude of the roughness,
$\lambda=\lambda_x=\lambda_y$ is the wavelength, 
$h$ the channel half height or pipe radius, 
$\Lambda = (4/\pi) k/\lambda_x$ denotes the solidity,
$L_x$ and $L_y$ are the streamwise and spanwise domain lengths,
while $N_x$, $N_y$, and $N_z$ are the number of cells in the streamwise, spanwise and wall-normal directions, respectively.
The pipe has an `O-grid' mesh so the product of $N_y\times N_z$ is given \citep{Chan15}.
$N_{\lambda}$ is the number of streamwise cells used to define a single sinusoidal roughness element,
$\Delta x^+$ and $\Delta y^+$ are the streamwise and spanwise grid spacings, and
$\Delta z_w^+$ and $\Delta z_h^+$ are the wall-normal grid spacings at the wall and at the channel centre, respectively.
$U_b^+$ is the bulk velocity through the channel. Note that this value for the minimal-span channel is larger than a full-span channel due to the altered outer layer of the flow.
Roughness function $\Delta U^+$ is calculated from the mean difference between smooth- and rough-wall flows, averaged over $40\le z^+\le z_c^+$.
Smooth-wall simulations do not have entries for roughness parameters.
}
\label{tab:sims}
\end{table}

\setlength{\unitlength}{1cm}
\begin{figure}
\centering
 \captionsetup[subfigure]{labelformat=empty}

	\includegraphics[]{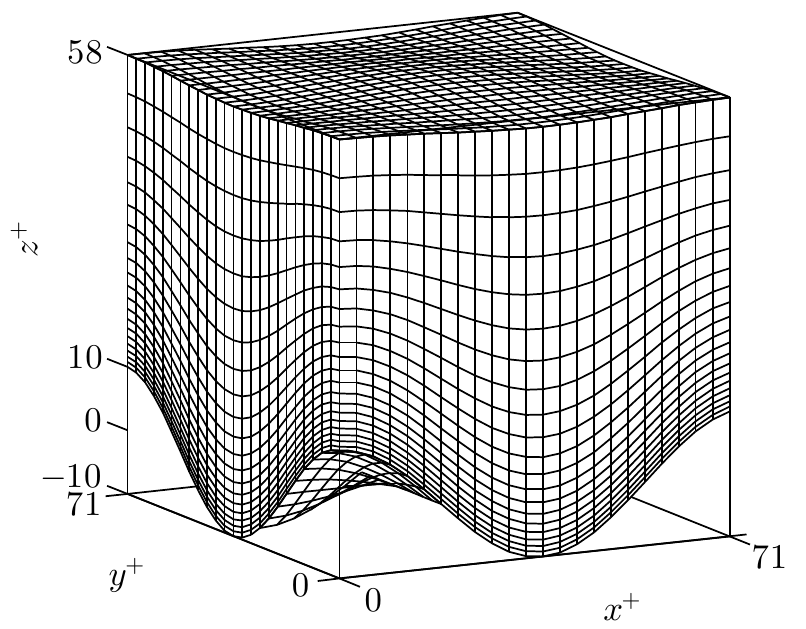}
	\vspace{-0.5\baselineskip}
\caption{Body-fitted mesh for a single roughness element with $\lambda^+ = 70.7$ and $k^+=10$ ($\Lambda = 0.18$), in the near-wall region of a channel at $Re_\tau=180$. Only every second wall-normal node is shown.}
	\label{fig:mesh}
\end{figure}

Table \ref{tab:sims} details the different simulations performed. The first set of simulations are mostly taken from \cite{Chan15} and are for full-span pipe flow, although there are some additional simulations that were not reported in the original paper ($\lambda^+=94$ and $\lambda^+=47$). These additional cases are performed with the same code and mesh structure. Some full-span channel simulations are detailed next, followed by the minimal-span channel cases. The final minimal-span channel case at $Re_\tau=180$ has a smooth-wall located at the crest of the roughness, which will be discussed later as representative of the limit of $\Lambda\rightarrow\infty$.
In the full-span pipe cases, an `O-grid' mesh is used so that the centre of the pipe has  a square-based mesh (\citealt{Chan15}). This means that the number of cells in the azimuthal--radial plane is given instead of the individual number of cells in the spanwise and wall-normal directions. In the channel flow cases, the spanwise and streamwise grid spacing is uniform. 
A hyperbolic tangent grid stretching is used in the wall-normal direction as in \cite{Moin82}, but with the stretching parameter $a=0.97$.
 For the rough-wall cases, a body-fitted grid is used to represent the sinusoidal roughness (figure \ref{fig:mesh}). Following \cite{Chan15}, the origin of the rough-wall channels is set as the mean roughness height, which for sinusoidal roughness is the same location as the smooth wall origin, $\varepsilon^+=0$. This coincides with the hydraulic origin, which ensures a collapse of the total stress profile (sum of the Reynolds shear stress and viscous stress) in the outer layer for smooth- and rough-wall flows.

The spanwise domain length $L_y$ of the minimal-span channels are set according to the guidelines recommended in \cite{Chung15}, namely that $L_y^+>100$, $L_y>\lambda_y$, and $L_y > k/0.4$. The first constraint ensures the near-wall cycle is resolved by the minimal channel. The second constraint is to ensure that at least one repeating unit of roughness is represented, while the final constraint is required to ensure that the roughness is submerged in fully resolved turbulent flow. Note that the final constraint may need to be larger, e.g., $L_y>3k/0.4$ for some roughness geometries if the roughness sublayer is larger. For the current sinusoidal roughness this appears to be unnecessary.  The streamwise domain length $L_x$ is usually over 2000 wall units, with some of the dense roughness simulations performed with $L_x^+\approx 1100$ owing to the high cost of the grid. A comparison between the two streamwise domain lengths is conducted for a rough wall with $\lambda^+ = 47.1$. The roughness function agrees within 4.5\% (table \ref{tab:sims}), indicating it will not influence the subsequent analysis or conclusions of this work.

The simulations are conducted at a friction Reynolds number of $Re_\tau=180$.  In general, the roughness function depends on the roughness and friction Reynolds numbers, $\Delta U^+(k^+,Re_\tau)$, with $\Delta U^+(k^+,Re_\tau)\rightarrow \Delta U^+ (k^+)$ as $Re_\tau\rightarrow \infty$. \cite{Chan15} showed that $\Delta U^+$ becomes  
independent of $Re_\tau$ for $Re_\tau\gtrsim 360$,
with the roughness function at $Re_\tau=180$ being approximately $0.6$ friction velocities higher than at $Re_\tau=360$ for roughness with matched viscous units (i.e., same $k^+$ and $\lambda^+$). The current data support this observation (table \ref{tab:sims}), with a simulation of roughness with $k^+=10$ and $\lambda^+=113$ at $Re_\tau=180$ having a roughness function approximately $0.4$ friction velocities larger than that of the same viscously scaled roughness at $Re_\tau=360$. 
Hence, if all the simulations were conducted at a higher friction Reynolds number but with matched $k^+$ values, there would only be a slight bulk shift in the roughness function for all roughness geometries but otherwise all results would  be similar to the $Re_\tau=180$ simulations. 
Moreover, this issue of friction Reynolds number becomes less important when using a minimal-span channel. The value of $Re_\tau = h U_\tau/\nu$ is not as relevant, as the largest eddies are restricted by the narrow spanwise width of the channel, rather than the channel half-height $h$. 

The simulations all have nominally the same $Re_\tau$ to minimise the influence of Reynolds number effects.
Minimal channels at low friction Reynolds numbers have been shown to have an $Re_\tau$ trend for the spanwise and wall-normal velocity fluctuations \citep{Hwang13},
and as mentioned above $\Delta U^+$ depends on $Re_\tau$ for low friction Reynolds numbers.
By having matched $Re_\tau$ we can ensure that the differences between smooth- and rough-wall flows are due to the roughness alone, and not Reynolds number.
This means the total drag force exerted on the wall is the same for all cases, however it will be shown that the partition between viscous and pressure drag components will vary.
   Furthermore, while the drag force is matched for all cases, different rough-wall flows cannot sustain the same mass flux and so the bulk velocity $U_b^+$ will vary depending on the roughness topography (table \ref{tab:sims}). 
   The rough wall cannot sustain the same flow rate as a smooth wall, so the mean velocity profile is shifted down by an amount $\Delta U^+$. This can then be related to the skin-friction coefficients through (\ref{eqn:DU_CF}), where this equation only holds if $Re_\tau$ is matched.


\section{Minimal-span rough-wall channel}
\label{sect:minspanvalid}
\setlength{\unitlength}{1cm}
\begin{figure}
\centering
 \captionsetup[subfigure]{labelformat=empty}

	\subfloat[]{
		\includegraphics[width=0.49\textwidth]{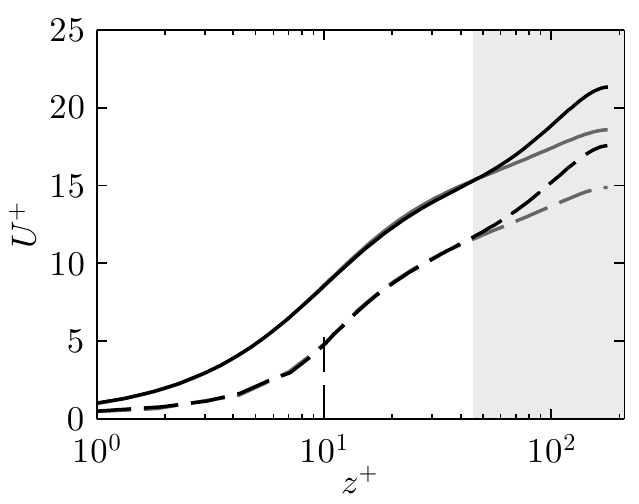}
		\label{fig:UxValid1}
	}
	\subfloat[]{
		\includegraphics[width=0.49\textwidth]{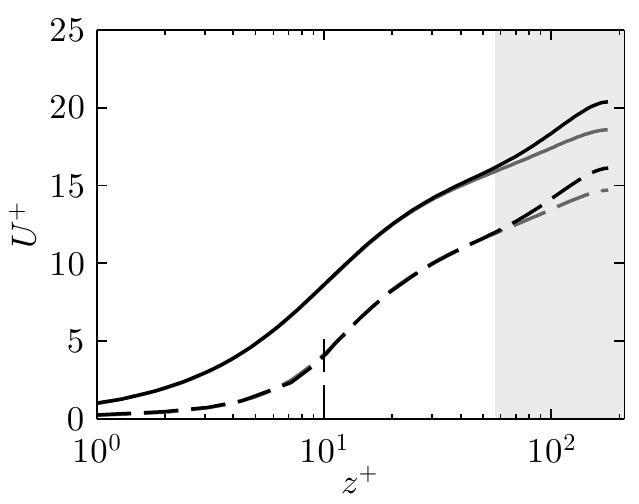}
		\label{fig:UxValid2}
	}
	\put(-12.35,2.6){\includegraphics{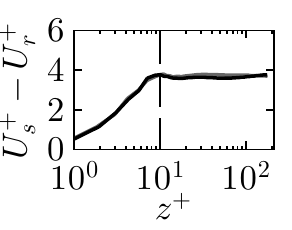}}
	\put(-13.65,4.8){(\emph{a})}
	\put(-5.5,2.6){\includegraphics{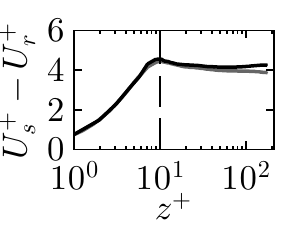}}
	\put(-6.8,4.8){(\emph{b})}
	\put(-10.15,1.1){$k^+$=$10$}
	\put(-8.66,1){\line(0,1){0.5}}
	\put(-8.6,1.1){$z_c^+$=$0.4L_y^+$}
	\put(-3.3,1.1){$k^+$=$10$}
	\put(-1.60,1){\line(0,1){.5}}
	\put(-1.52,1.1){$z_c^+$=$0.4L_y^+$}
	\vspace{-2.2\baselineskip}
\caption{Mean streamwise velocity profile for (\emph{a}) $\Lambda = 0.11$, $L_y^+=113$ and for (\emph{b}) $\Lambda=0.18$, $L_y^+=141$. Line styles are:
solid, smooth wall;
dashed, rough wall;
grey, full span;
black, minimal span.
Vertical dashed line shows crest of roughness ($k^+=10$).
Light grey shading starts at $z_c^+=0.4 L_y^+$, the point where the minimal-span channel deviates from the full-span \citep{Chung15}.
The inset shows  the difference in velocity from the smooth- and rough-wall simulations, $U_s^+ - U_r^+$.}
	\label{fig:velProfValid}
\end{figure}
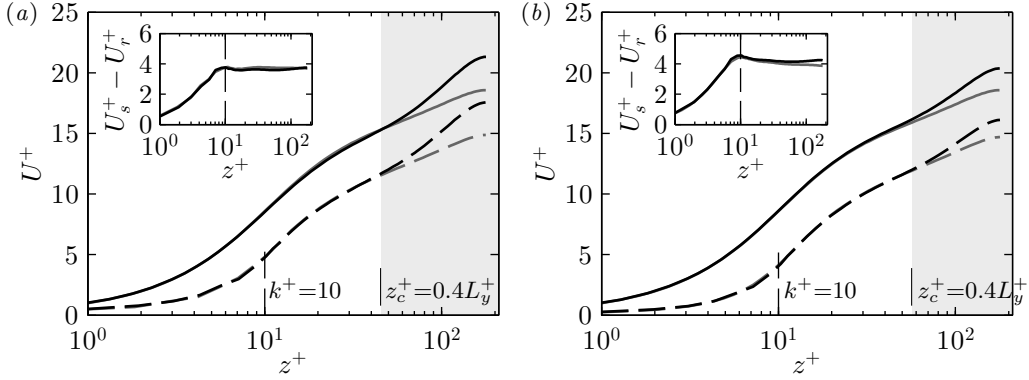

Validation of the minimal-span channel is first performed to show that the near-wall behaviour is accurately captured in both the sparse and dense roughness regimes. The mean velocity profile is shown in figure \ref{fig:velProfValid} for the full-span and minimal-span channel simulations of cases with roughness wavelengths of 113 and 71 wall units. These simulations have solidity values of 0.11 and 0.18, which place them in the sparse and dense regimes, respectively. The mean velocity profile of the minimal-span channel shows good agreement with the full domain, up until $z^+\approx50$, at which point the minimal-span channel has an increased streamwise velocity. \cite{Flores10} and \cite{Hwang13} determined the critical wall-normal location where the streamwise mean velocity of the minimal-span channel deviates from the full-span channel as $z_c\approx 0.3L_y^+$, while a scaling of $z_c^+\approx0.4 L_y^+$ was suggested by \cite{Chung15}.  The grey shaded regions in \mbox{figure \ref{fig:velProfValid}} indicates the critical wall-normal position $z_c^+=0.4L_y^+$, which is seen to be consistent with the current results. Above this location the minimal-span channel cannot represent the full scale of turbulent structures in the outer layer, resulting in the velocity increasing compared to the full-span channel. Following \cite{Flores10}, we will refer to the flow below $z_c$ as being `healthy' turbulence, as it is the same as the full-span channel.
In the dense roughness case (figure \ref{fig:velProfValid}\emph{b}) the spanwise domain is larger ($L_y^+ = 141\Rightarrow z_c^+\approx56$), so a larger proportion of the turbulent scales are captured and the velocity does not increase as significantly as with the sparse roughness (figure \ref{fig:velProfValid}\emph{a}), which has a smaller spanwise domain ($L_y^+ = 113\Rightarrow z_c^+\approx 45$).  Looking at the difference in velocity between smooth- and rough-wall flows (insets of figure \ref{fig:velProfValid}), there is good agreement between the minimal- and full-span channels. There is a small difference of around 5\% for the dense roughness case (inset of figure \ref{fig:velProfValid}\emph{b}) for $z^+\gtrsim 60$, which is likely due to the slightly different grid that is used (see $\Delta z_h^+$ in table \ref{tab:sims}).

The sparse roughness case ($\Lambda = 0.11$) has an identical velocity profile to the smooth-wall flow (except for the offset $\Delta U^+$) from just above the crest of the roughness to the channel centre. This is seen by the velocity difference remaining  approximately constant for $z^+ \gtrsim k^++5$ in the inset of figure \ref{fig:velProfValid}(\emph{a}). Here, the roughness function $\Delta U^+\approx3.7$, varying by only $\pm 1\%$ over $k^++5<z^+<180$.
For the dense roughness case ($\Lambda = 0.18$), however, the velocity difference only reaches a constant above $z^+\approx k^++25$, with a roughness function of $\Delta U^+=3.9$, varying by only $\pm3\%$ over $k^++25<z^+<180$. Given that the dense regime shows a changing velocity profile in the near-wall region, the value for $\Delta U^+$ is obtained by the average value of $U_s^+-U_r^+$ over $40\leq z^+\leq z_c^+$. For full-span channels $z_c^+=h^+$ while for minimal-span channels $z_c^+= 0.4L_y^+$ \citep{Chung15}.

\setlength{\unitlength}{1cm}
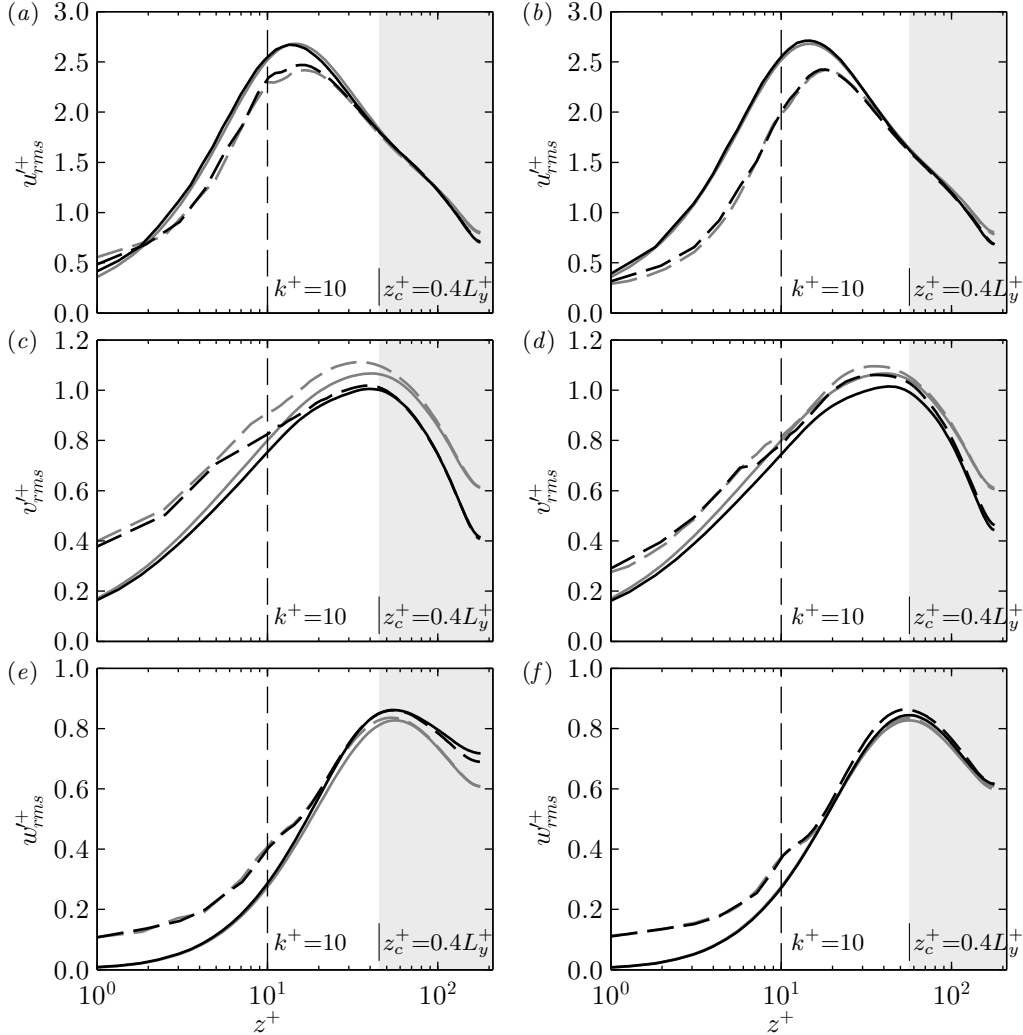
\begin{figure}
\centering
 \captionsetup[subfigure]{labelformat=empty}

	\subfloat[]{
		\includegraphics[trim=0 2 0 5,clip=true,width=0.49\textwidth]{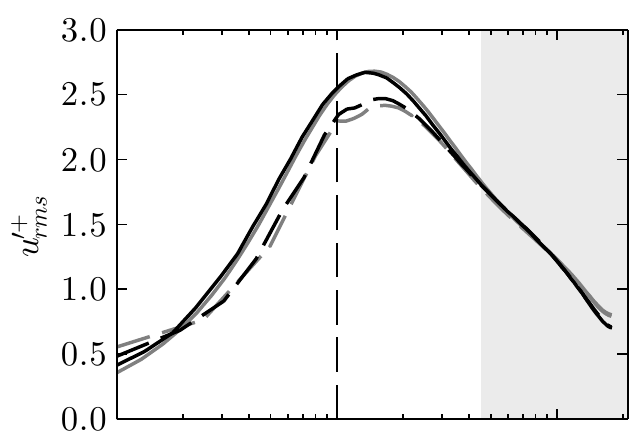}
	}
	\subfloat[]{
		\includegraphics[trim=0 2 0 5,clip=true,width=0.49\textwidth]{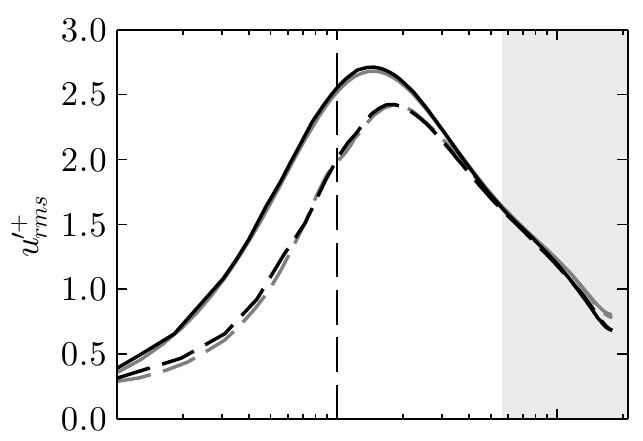}
	}
	\put(-13.55,4.05){(\emph{a})}
	\put(-6.7,4.05){(\emph{b})}
	\put(-10,0.38){$k^+$$=$$10$}
	\put(-8.6,0.25){\line(0,1){0.5}}
	\put(-8.55,0.38){$z_c^+$=$0.4L_y^+$}
	\put(-3.12,0.38){$k^+$$=$$10$}
	\put(-1.55,0.25){\line(0,1){.5}}
	\put(-1.50,0.38){$z_c^+$=$0.4L_y^+$}
	\\
	\vspace{-0.8cm}
	\subfloat[]{
		\includegraphics[trim=0 2 0 5,clip=true,width=0.49\textwidth]{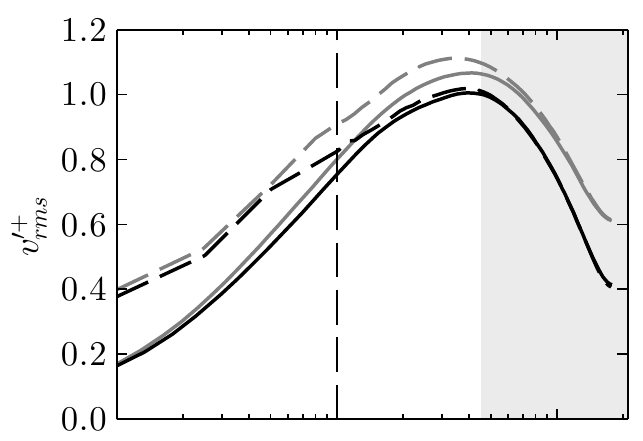}
	}
	\subfloat[]{
		\includegraphics[trim=0 2 0 5,clip=true,width=0.49\textwidth]{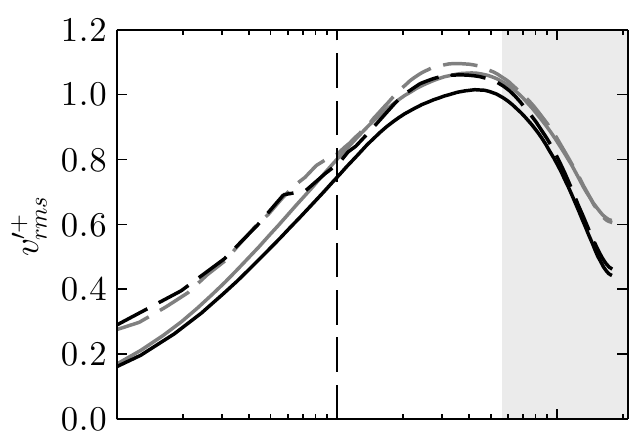}
	}
	\put(-13.55,4.05){(\emph{c})}
	\put(-6.7,4.05){(\emph{d})}
	\put(-10,0.38){$k^+$$=$$10$}
	\put(-8.6,0.25){\line(0,1){0.5}}
	\put(-8.55,0.38){$z_c^+$=$0.4L_y^+$}
	\put(-3.12,0.38){$k^+$$=$$10$}
	\put(-1.55,0.25){\line(0,1){.5}}
	\put(-1.50,0.38){$z_c^+$=$0.4L_y^+$}
	\\
	\vspace{-0.8cm}
	\subfloat[]{
		\includegraphics[trim=0 2 0 5,clip=true,width=0.49\textwidth]{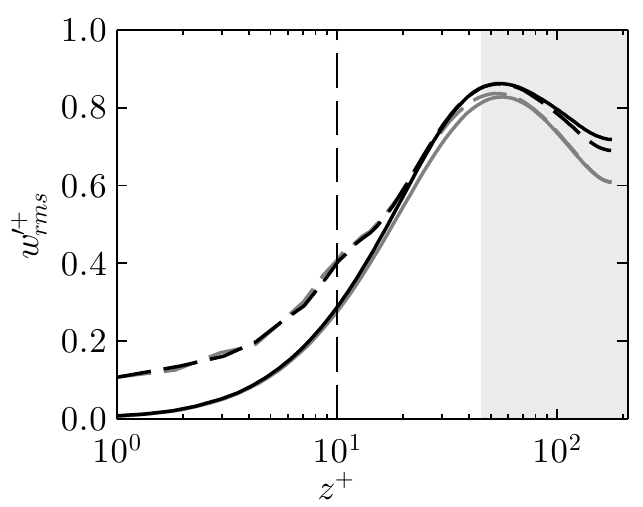}
	}
	\subfloat[]{
		\includegraphics[trim=0 2 0 5,clip=true,width=0.49\textwidth]{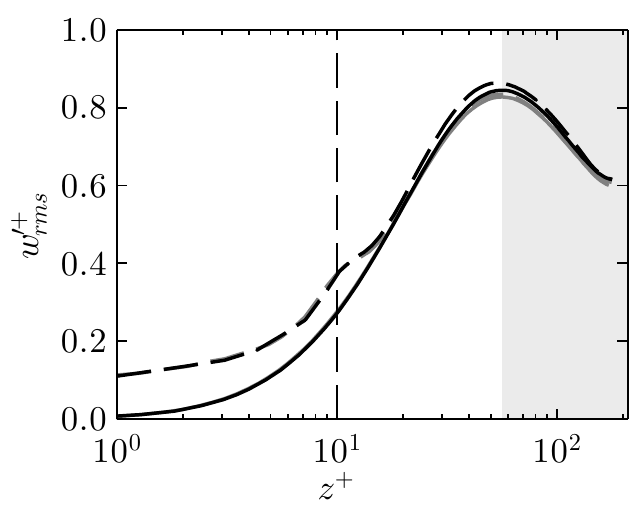}
	}
	\put(-13.55,4.7){(\emph{e})}
	\put(-6.7,4.7){(\emph{f})}
	\put(-10,1.1){$k^+$$=$$10$}
	\put(-8.6,0.95){\line(0,1){0.5}}
	\put(-8.55,1.1){$z_c^+$=$0.4L_y^+$}
	\put(-3.12,1.1){$k^+$$=$$10$}
	\put(-1.55,0.95){\line(0,1){.5}}
	\put(-1.50,1.1){$z_c^+$=$0.4L_y^+$}
	\vspace{-0.75cm}
\caption{Root-mean-square velocity fluctuations in the (\emph{a},\emph{b}) streamwise, (\emph{c},\emph{d}) spanwise, and  (\emph{e},\emph{f}) wall-normal directions for (\emph{a},\emph{c},\emph{e}) $\Lambda=0.11$, $L_y^+=113$ (sparse roughness) and (\emph{b},\emph{d},\emph{f}) $\Lambda = 0.18$, $L_y^+=141$ (dense roughness). Line styles are the same as figure \ref{fig:velProfValid}.
}
	\label{fig:velFlucValid}
\end{figure}

The root-mean-square velocity fluctuations of all three velocity components are shown in figure \ref{fig:velFlucValid}. Firstly we will look at the smooth-wall data (solid lines) for full- and minimal-span channels. The data show qualitatively similar behaviour to the smooth-wall minimal-span channels of \cite{Hwang13} at the same friction Reynolds number.  Here, \citeauthor{Hwang13} showed that towards the channel centreline, the minimal-span channel displayed  reduced streamwise and spanwise velocity fluctuations and enhanced wall-normal velocity fluctuations when compared to full-span channel flow.
It was also observed that the spanwise velocity fluctuations in the minimal channel begin to deviate from the full-span channel immediately at the wall, while the streamwise and wall-normal velocity fluctuations are in agreement in the near-wall region ($z^+ \lesssim 30$). Evidently the spanwise velocity fluctuations in the near-wall region are more dependent on the spanwise domain length than are the streamwise and wall-normal velocity fluctuations.  However, the difference between the minimal-span and full-span spanwise velocity fluctuations is still only less than approximately 10\%, indicating that the spanwise motions that are not captured by the minimal-span channel (i.e., those larger than $\lambda_y>L_y$) do not contribute a significant amount of energy.

The velocity fluctuations of the minimal- and full-span rough-wall channels (dashed lines) are seen to collapse with their respective smooth-wall domains for large values of $z^+$. In the full-span simulations, this is predicted by Townsend's outer-layer similarity hypothesis \citep{Townsend76} which states that the flow only depends on the position above the wall for a domain with an infinitely large span. For example, the velocity defect is given by $(U_h-U(z))/U_\tau = f(z/h)$ and the root-mean-square velocity fluctuations by $u_{rms}/U_\tau=g(z/h)$ for $L_y^+\rightarrow \infty$ and $Re_\tau\rightarrow\infty$ . Importantly, it does not matter if the channel wall is rough or smooth. In full-span simulations the spanwise width is of the order of $L_y\gtrsim3h$ and use a finite $Re_\tau$, however several rough-wall studies  have shown that the hypothesis still holds, provided $L_y^+$ and $Re_\tau$ are matched between the smooth- and rough-wall flows (see, for example, \cite{Schultz09, Efros11, Chan15} and references therein). The current minimal-span simulations shows the hypothesis still holds in a more general sense wherein the spanwise domain width is simply restricted further than in full-span simulations, to the order of $L_y^+\approx 100$. The collapse for large $z^+$ shows that the general application of the hypothesis still holds, even in the case of the modified outer-layer flow that the minimal-span channel produces, provided $Re_\tau$ and $L_y^+$ are matched.

In the near-wall region, the streamwise (figure \ref{fig:velFlucValid}\emph{a},\emph{b}) and wall-normal (figure \ref{fig:velFlucValid}\emph{e},\emph{f}) velocity fluctuations of the rough-wall minimal-span channel are seen to be in very good agreement with the full-span roughness data, for both roughness wavelengths. As with the smooth-wall data, the velocity fluctuations only begin to deviate above $z^+ \gtrsim 30$--$40$. This is slightly below the point where the mean velocity profile of the minimal-span channel deviates from that of the full-span channel ($z_c^+\approx 0.4L_y$), although not substantially different. The streamwise and wall-normal velocity fluctuations could be argued to obey these scalings, although further simulations in which the spanwise domain length is varied would be required to definitively answer this question. However, it seems reasonable that the departure of the velocity fluctuations would also scale on $L_y$. There is also a dependence on $Re_\tau$ as discussed in \cite{Hwang13}, in that the large outer-layer motions that become increasingly dominant in higher $Re_\tau$ flows are not captured by the minimal-span channel. The minimal-span rough-wall spanwise velocity fluctuations (figure \ref{fig:velFlucValid}\emph{c},\emph{d}) perform similarly to that of the minimal-span smooth-wall channel in that they are slightly damped compared to the full span channel, even close to the wall. The ability to reproduce velocity fluctuations within the roughness elements is desirable in geophysical applications; for example, pollutant dispersion within urban or natural canopies \citep{Macdonald00}. Attempting to obtain reliable experimental data within this region is difficult and costly, while it is  straightforward with the minimal-span channel.

\setlength{\unitlength}{1cm}
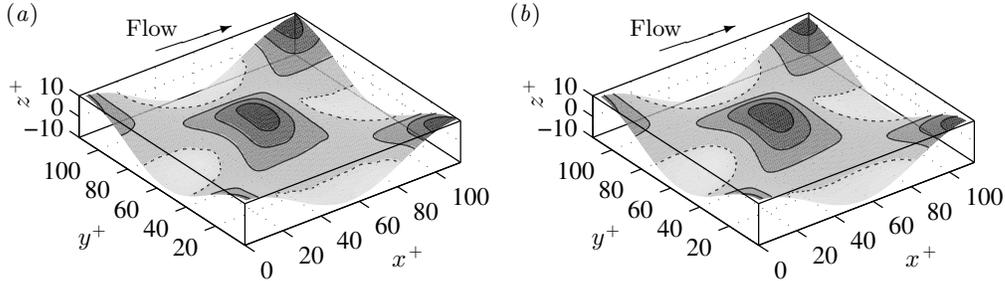
\begin{figure}
\centering
 \captionsetup[subfigure]{labelformat=empty}

	\subfloat[]{
		\includegraphics[width=0.49\textwidth]{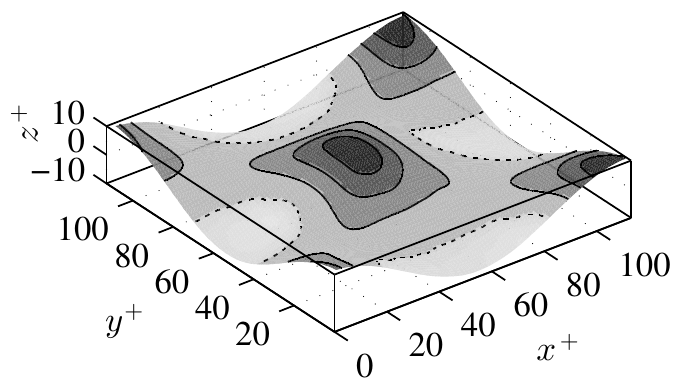}
	}
	\subfloat[]{
		\includegraphics[width=0.49\textwidth]{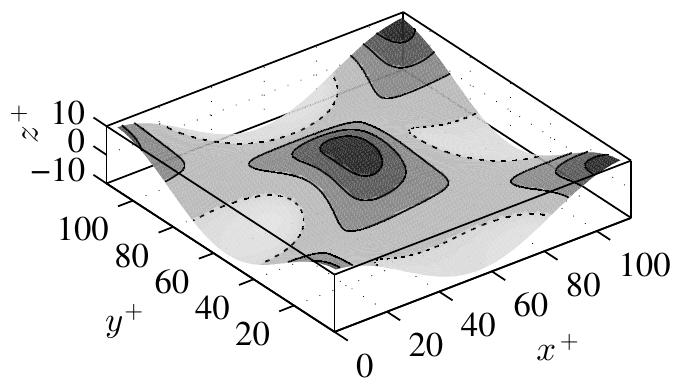}
	}
	\put(-13.5,3.5){(\emph{a})}
	\put(-6.8,3.5){(\emph{b})}
	\put(-11.5,3.1){\vector(3,1){1}}
	\put(-11.9,3.3){Flow}
	\put(-4.8,3.1){\vector(3,1){1}}
	\put(-5.2,3.3){Flow}
	\vspace{-2.2\baselineskip}
\caption{Streamwise component of the viscous stress $\tau_{w,x}^+$ in the (\emph{a}) full-span and (\emph{b}) minimal-span channels, for a rough wall with wavelength $\lambda^+=113$ ($\Lambda=0.11$).
Dotted line shows zero stress (recirculation) region, contour lines show $\tau_{w,x}^+=(0.8,1.6,2.4)$.
}
	\label{fig:spatialDragValid}
\end{figure}

The streamwise component of viscous stress, $\tau_{w,x}^+$, is shown in figure \ref{fig:spatialDragValid} for the full- and minimal-span channels for roughness with a solidity of $\Lambda = 0.11$. The minimal-span channel is accurately reproducing the spatial variations of viscous stress that are seen in the full-span channel. The pressure drag (not shown) is also faithfully reproduced in the minimal-span channel, with the ratio of pressure to total (pressure + viscous) drag being 0.37 for this roughness.

Given that comparisons made between the minimal- and full-span channels, for both sparse and dense roughness regimes, show good agreement with each other, the remainder of the study will primarily deal with minimal-span channel results.


\section{Dense roughness limit}
\label{sect:dense}
\setlength{\unitlength}{1cm}
\begin{figure}
\centering
 \captionsetup[subfigure]{labelformat=empty}

	\includegraphics[]{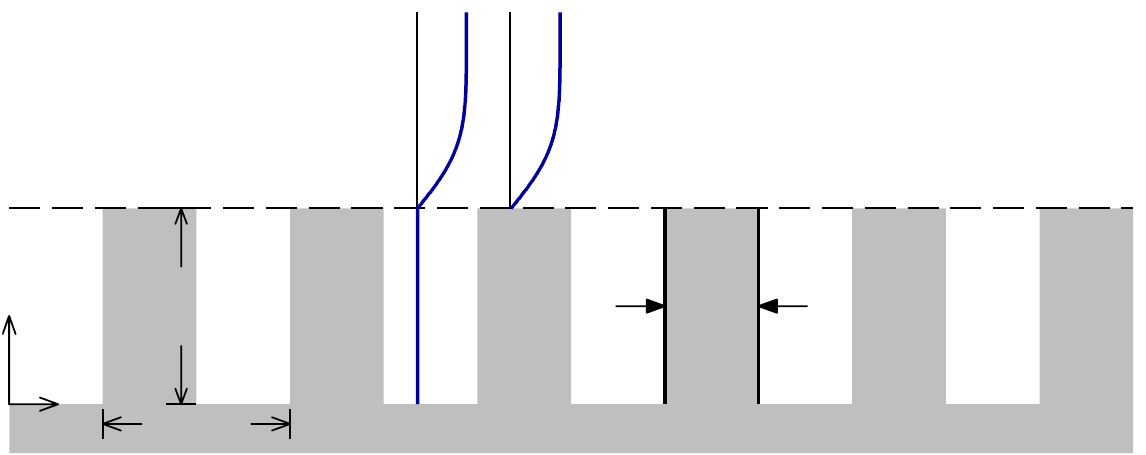}
	\put(-10.2,0.2){$\lambda\ll \frac{\nu}{U_\tau}$}
	\put(-11.0,0.6){$x$}
	\put(-11.45,1.3){$z$}
	\put(-10.2,1.4){$k>\frac{\nu}{U_\tau}$}
	\put(-5.8,3.7){$U(z)$}
	\put(-5.4,1.2){$f_1$}
	\put(-3.6,1.2){$f_2$}
	\vspace{-0.5\baselineskip}
\caption{Representation of the  limit of $\Lambda\rightarrow\infty$. Assuming zero velocity below roughness crest, implying an offset smooth-wall at the roughness crest, $z_w=k$ (dashed line).}
	\label{fig:solLimSketch}
\end{figure}
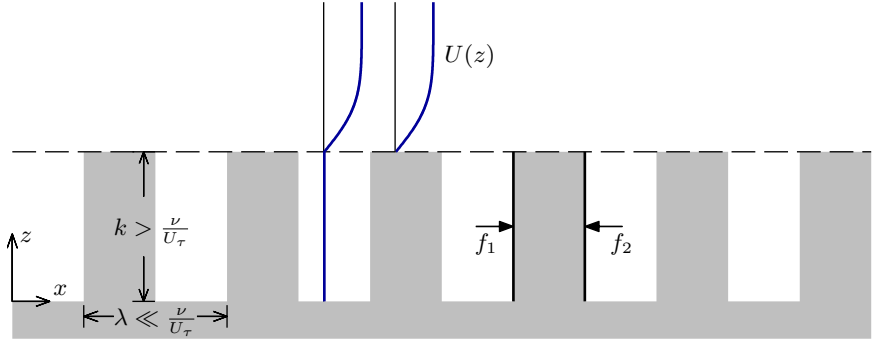

If we consider the limit of solidity $\Lambda \rightarrow \infty$, then for sinusoidal roughness the wavelength is tending to zero and so it is much smaller than the viscous length scale. 
This small-scale roughness could conceptually be thought of as a bed of tightly packed pins, and the exceedingly small roughness length scale would limit any flow occurring within the roughness canopy. There would be some flow within the roughness canopy due to the pressure drop down the channel, similar to flow within a porous medium. However, 
a subsequent Stokes flow analysis suggests that the flow within the roughness canopy becomes negligible relative to the flow above the roughness. In this case, the wall-parallel plane of the roughness crests has a velocity that can be assumed to be zero. This is then equivalent to a smooth wall, located at the roughness crests, $z_w^+=k^+$.
For the current roughness, this can be thought of as an $Re_\tau=180$ rough-wall channel with the origin at $z_0^+=0$ and a solid block of roughness with $k^+=10$ (table \ref{tab:sims}). Equivalently, it can be 
regarded as an $Re_\tau=170$ smooth-wall channel with an offset of $10$ wall units in the wall-normal position.  This limiting case of an $Re_\tau=170$ channel is simulated and the roughness function is simply the difference in centreline velocity between smooth-wall flow at $Re_\tau=180$ and at $Re_\tau=180-k^+$, which has a value of $\Delta U^+ = 0.93$.

 It is also desirable to know the ratio of pressure drag to total drag acting on the roughness. This can be obtained by considering the pressure drag across a single roughness element (figure \ref{fig:solLimSketch}), $f_p = f_1-f_2 = p_1k\lambda_y - p_2k\lambda_y$, where $p_1$ and $p_2$ are the pressures acting on the front and rear faces,
  respectively,  and $\lambda_y$ is the spanwise extent of the element. 
  If there is negligible flow within the roughness cavities in the limit of $\Lambda\rightarrow\infty$, then the pressure drop across the element is simply the pressure gradient applied to the channel multiplied by the streamwise distance, $p_1-p_2 = -\dd p/\dd x \cdot\lambda/2$. The total pressure drag for all the elements would then just be the individual pressure drag for a single element multiplied by the number of elements, $F_p = f_p L_x/\lambda\cdot L_y/\lambda_y = -k L_x L_y\dd p/\dd x$. While the representation in figure \ref{fig:solLimSketch} shows rectangular roughness elements, the result is the same if considering sinusoidal roughness.
The viscous drag now has to balance the pressure gradient applied to the fluid volume not occupied by roughness, $F_\nu = -L_x \dd p/\dd x\cdot L_y (h-k)$. As such, in the limit of $\Lambda\rightarrow\infty$, the ratio $F_p/F_T = F_p/(F_p+F_\nu)=k/h$ and for the current roughness this is approximately $F_p/F_T\approx0.056$. By considering this limiting case, we can see that both the roughness function and pressure to total drag ratio in figure \ref{fig:solidity} must go on to decrease to become nearly zero for extremely dense roughness. A subsequent analysis will suggest that the limit of $\Lambda\rightarrow\infty$ is reached when $\Lambda\approx4.2\Rightarrow\lambda^+\approx3.0$.

This argument of the limit of $\Lambda\rightarrow\infty$ being represented by an offset smooth wall at the roughness crest is only applicable to internal flows. $\Delta U^+$ can be seen to be tending to a finite number for $\Lambda\rightarrow\infty$ and this is due to continuity, as the blockage from this offset smooth-wall requires a reduced mass flux through the channel. It also follows that this limit is proportional to $k/h$. However, external flows do not suffer from this blockage effect, meaning that the limit of $\Lambda\rightarrow\infty$ will result in $\Delta U^+\rightarrow 0$ for boundary layers. This represents a fundamental difference between internal and external flows.

\setlength{\unitlength}{1cm}
\begin{figure}
\centering
 \captionsetup[subfigure]{labelformat=empty}

	\subfloat[]{
		\includegraphics[width=0.49\textwidth]{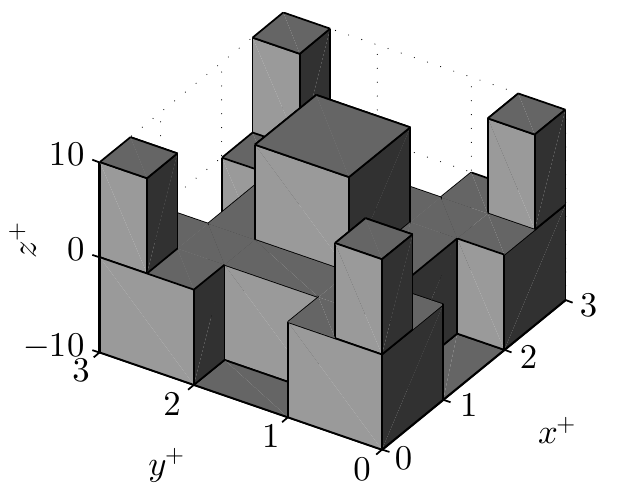}
	}
	\subfloat[]{
		\includegraphics[width=0.49\textwidth]{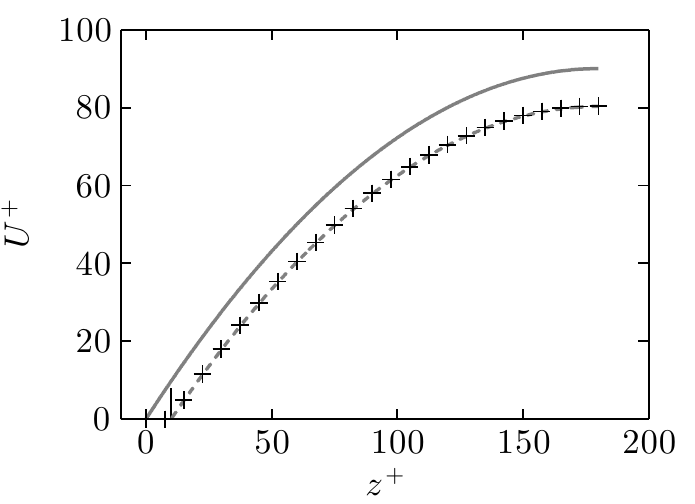}
	}
	\vspace{-2.0\baselineskip}
	\put(-13.55,4.85){(\emph{a})}
	\put(-6.75,4.85){(\emph{b})}
\caption{
(\emph{a}) Stokes flow over block roughness where $k/\lambda=3.33$. Note the $z$-axis is not to scale.
(\emph{b}) Stokes flow velocity profile. Line styles:
\protect\raisebox{0.4ex}{\protect\scalebox{1.0}{$\color{black}\boldsymbol{\pmb{+}}$}}, Stokes flow data over the cuboid blocks of (\emph{a});
\protect\raisebox{0.8ex}{\color{LGrey}\linethickness{0.5mm}\line(1,0){0.6}}, analytic Stokes flow with smooth wall at $z=0$ (\ref{eqn:lamFlow});
\protect\raisebox{0.8ex}{\color{LGrey}\linethickness{0.5mm}\line(1,0){0.2}\hspace{0.15cm}\line(1,0){0.2}}, analytic Stokes flow with offset smooth wall at $z=k$ (\ref{eqn:offsetLamFlow}).
Roughness crest is at $z^+=10$.
}
	\label{fig:stokesVel}
\end{figure}

The wavelength of the roughness, $\lambda$, would tend to zero as $\Lambda\rightarrow\infty$. When it is less than the viscous length scale, it would limit any inertial flow behaviour in the near-wall region resulting in Stokes flow occurring around the roughness elements. This Stokes flow regime can be simulated by neglecting the non-linear and time-dependent terms in the Navier--Stokes equations, so that only the viscous and pressure gradient terms remain. This would allow us to investigate the assumption that the limit of $\Lambda\rightarrow\infty$ can be modelled by an offset smooth wall, as the Stokes flow data can be compared with the classical parabolic velocity profile for smooth-wall flow.
To this end, Stokes flow is simulated over a rough surface with a large height-to-wavelength ratio of $k/\lambda=3.33$. As sinusoidal roughness would be prohibitively expensive to grid, cuboid blocks are used with a similar layout to the three-dimensional sinusoidal roughness (figure \ref{fig:stokesVel}\emph{a}).
 The cuboid roughness simulated here has a solidity of $\Lambda=4.44$ and the blockage ratio $k/h=1/18$ is the same as the turbulence simulations.
 
A single repeating unit is simulated with periodic boundary conditions in the streamwise and spanwise directions, and a slip wall is positioned at $z=h$.
The streamwise and spanwise directions are discretised with 96 evenly spaced cells, while between the roughness trough and crest the wall-normal direction has 64 evenly spaced cells. Above the roughness crest, there are 64 cells with a cell-to-cell expansion ratio of 1.06. An additional simulation was performed with twice as many cells in each direction, which showed negligible differences in the velocity profile.

The mean velocity of these Stokes flow simulations is shown in figure \ref{fig:stokesVel}(\emph{b}) along with the expected parabolic profiles for Stokes flow in a channel. This is done for both a standard smooth-wall flow where the wall is located at $z=0$,
  \begin{equation}
  U = -\frac{1}{2\mu}\df{p}{x}\left(h^2-(h-z)^2\right),
  \label{eqn:lamFlow}
  \end{equation}
  and for the offset smooth-wall at $z=k$ which is representative of the limit of $\Lambda\rightarrow\infty$,
    \begin{equation}
  U = -\frac{1}{2\mu}\df{p}{x}\left((h-k)^2-(h-z)^2\right).
  \label{eqn:offsetLamFlow}
  \end{equation}
  In this case, the channel half-height is effectively reduced to $h-k$ while maintaining the same dynamic viscosity $\mu=\nu/\rho$ and same pressure drop. Note that in this paper we define the origin of $z$ at the wall, whereas most textbook derivations of (\ref{eqn:lamFlow}) define the origin at the channel centre.
As can be seen in figure \ref{fig:stokesVel}(\emph{b}), there is excellent agreement between the simulation and the offset smooth wall velocity (\ref{eqn:offsetLamFlow}). 

The pressure drop applied to the channel only results in a weak velocity through the roughness canopy. This weak flow is similar to flow through a porous media, which can be predicted through Darcy's law, $U = -\kappa_p/(\mu) \mathrm{d}p/\mathrm{d}x$, where $\kappa_p$ is the intrinsic permeability. We can use the empirical relationship $\kappa_p/d^2=6.17\times10^{-4}$ \citep{Krumbein43}, where $d$ is the pore diameter, taken here to be the length scale of the top roughness cuboids in figure \ref{fig:stokesVel}(\emph{a}), $d=\lambda/3$. The predicted roughness canopy velocity  for the current flow conditions would then be $U^+\approx6\times10^{-4}$, which is not dissimilar to the actual velocity of $U^+\approx 4\times10^{-4}$.  The slip velocity at the roughness crest is $U(z=k)^+=0.3$, but the agreement with the offset smooth wall profile suggests that this does not have a significant effect on the flow.  This indicates the flow through the roughness canopy for small wavelengths is similar to a porous media flow, and that the velocity itself is negligible relative to the flow above.
This idea is consistent with the classical result of \cite{Beavers67}, who proposed that the roughness crest slip velocity is proportional to $\sqrt{\kappa_p}$, for large $h/\sqrt{\kappa_p}$. For the scalings above of $\sqrt{\kappa_p}\sim\lambda$, this supports the argument that $U(z=k)\rightarrow0$ as $\lambda^+\rightarrow0$.
 Hence, Stokes flow over roughness with a large  solidity can be modelled as an offset smooth wall. As discussed above, the turbulent flow around roughness with $\lambda^+\rightarrow0$ will also be dominated by viscous processes, indicating this offset smooth-wall assumption will also apply to turbulent flows. 
Indeed, \cite{Rosti15} performed DNS of turbulent flow at $Re_\tau=180$ over porous walls and showed a very similar result in which the mean velocity profile resembled an offset smooth wall. 
 It is important to emphasise that by assuming Stokes flow conditions, we have assumed that $\lambda^+$ is small in a true turbulent flow. The value of $\lambda^+$ shown in figure \ref{fig:stokesVel}(\emph{a}) is therefore not indicative of when the limiting condition is reached, but rather 
 what the flow would look like in the limit. In this case, it informs us that when the flow is dominated by viscous processes it can indeed be modelled by an offset smooth wall.


\section{Rough-wall results}
\label{sect:rough}


\subsection{Mean statistics}

\setlength{\unitlength}{1cm}
\begin{figure}
\centering
 \captionsetup[subfigure]{labelformat=empty}

	\subfloat[]{
		\includegraphics[width=0.49\textwidth]{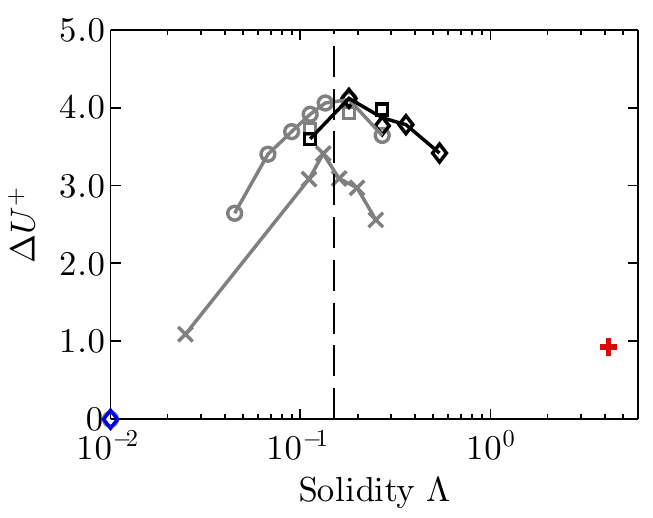}
	}
	\subfloat[]{
		\includegraphics[width=0.49\textwidth]{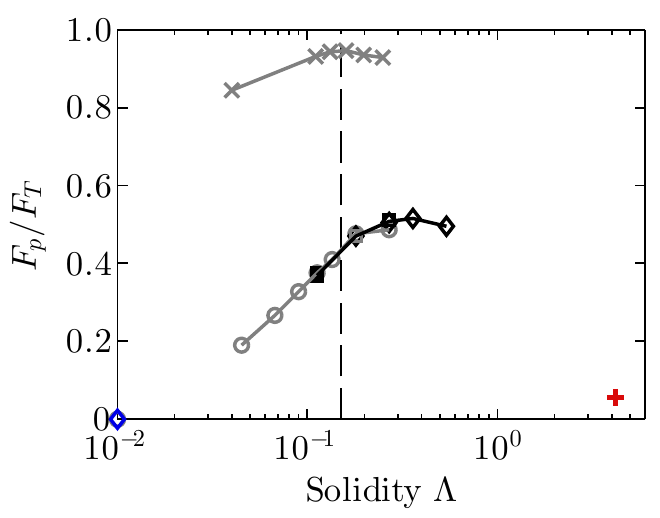}
	}
	\put(-13.6,4.85){(\emph{a})}
	\put(-6.7,4.85){(\emph{b})}
	\vspace{-2.2\baselineskip}
	\put(-12.1,4.5){Sparse}
	\put(-9.6,4.5){Dense}
	\put(-5.25,4.45){Sparse}
	\put(-2.7,4.45){Dense}
	\put(-12.35,1.1){$\Lambda=0$}
	\put(-5.45,1.1){$\Lambda=0$}
	\put(-8.2,1.35){$\Lambda\rightarrow\infty$}
	\put(-1.4,1.35){$\Lambda\rightarrow\infty$}
\caption{(Colour online)
(\emph{a}) Roughness function, $\Delta U^+$ and (\emph{b}) ratio of pressure to total drag force, $F_p/F_T$, against solidity $\Lambda$. 
Symbols:
\protect\raisebox{0.5ex}{\protect\scalebox{0.8}{$\bigcirc$}}, full-span pipe;
$\square$, channel with $L_x^+=2262$;
$\diamondsuit$, channel with $L_x^+=1131$.
\protect\raisebox{0.4ex}{\protect\scalebox{1.0}{$\color{myred}\boldsymbol{\pmb{+}}$}}, limit of $\Lambda\rightarrow\infty$.
Grey symbols: full-span pipe or channel.
Other colours: minimal-span channel.
Vertical dashed line at $\Lambda = 0.15$ demarcates the sparse and dense roughness regimes.
Smooth wall indicated by $\Lambda=0$.
Grey line with crosses indicate the cuboid rough surfaces from figure 5 of \cite{Leonardi10}, where the  roughness function is offset by $\Delta U^+-12$.
}
	\label{fig:solidity}
\end{figure}

To investigate the effect of roughness solidity, the roughness function $\Delta U^+$ is shown as a function of the frontal solidity, $\Lambda$, in figure \ref{fig:solidity}(\emph{a}). A vertical dashed line corresponding to $\Lambda = 0.15$ is shown, which is the point at which the dense regime is thought to begin \citep{Jimenez04,Flack14}, although recently a higher value of $\Lambda=0.21$ has been suggested \citep{Placidi15}. It can be seen that the current data appear to show the dense regime begins when the solidity is around $\Lambda \approx 0.18$, although the range of $0.15\leq\Lambda\leq0.2$ still appears to be a reasonable estimate for the point where the regimes change.  It is expected that the change in regime is also dependent on the roughness geometry; here we are using a regular arrangement of roughness elements in the transitionally rough regime. At any rate, a  reduction in the roughness function is observed for $\Lambda\geq0.18$ showing that the dense roughness regime has been obtained. 

The fully rough cuboid surfaces of \cite{Leonardi10} are also shown, although the roughness function is offset by 12 friction velocities. The roughness function is clearly maximum at $\Lambda\approx0.13$, and reduces more significantly in the dense regime than the current sinusoidal roughness. This is due to the approach used by \cite{Leonardi10}, whereby the cubes maintain the same dimensions but are moved progressively closer together.  As such, the total fluid-occupied volume of the channel reduces with increasing solidity, until at $\Lambda=1$ there is a smooth wall at the roughness crest. Hence, the roughness function $\Delta U^+\rightarrow0$ for $\Lambda=1$. In contrast, the current roughness maintains a constant fluid-occupied volume, so that this limit only occurs for $\Lambda\rightarrow\infty$. Nevertheless, it is interesting that the current transitionally rough flow exhibits such a similar trend to a fully rough flow, where the flow no longer depends on the roughness Reynolds number.
The current roughness function only varies over $2.1\le\Delta U^+\le4.1$, which is not as large as the range seen in studies going from the hydraulically smooth to the fully rough regimes where $0\le \Delta U^+\le O(10)$.
Despite the relatively narrow range of $\Delta U^+$ of this study, it is important to mention that the roughness function is not the only indicator of how much of an  effect a rough wall is having on the flow.
  \cite{Antonia01} showed that two different surfaces with nominally the same $\Delta U^+$ can have appreciable differences in the turbulence statistics. Drag-reducing riblets have fascinated engineers for decades, yet the roughness function varies by less than 1 friction velocity from a smooth wall \citep{Spalart11}. The fully rough data of \cite{Leonardi10} in figure \ref{fig:solidity}(\emph{a}) show that studies of sparse and dense roughness do not necessarily elicit large variations in $\Delta U^+$, however as will be shown there is still rich flow dynamics that can be analysed.

The differences between transitionally and fully rough flows become apparent when considering models for predicting $\Delta U^+$. The model of \cite{Macdonald98} predicts the equivalent sand grain roughness $k_s^+$ for fully rough flows, using the roughness drag coefficient $C_D$, solidity $\Lambda$, and zero-plane displacement height $d/k$. 
For the case of $\Lambda=0.18$, \cite{Chan15} determined $k_s/k=4.1$  yet the model of \cite{Macdonald98} with the current transitionally rough data overpredicts $k_s$ by 50\% to be $k_s/k\approx6.2$. This is for a drag coefficient calculated to be $C_D\approx1.3$ which is not dissimilar to  the value of $C_D=1.2$ recommended for fully rough cubes. However, the displacement height is taken to be $d/k=0$ as the rough-wall channel maintains the same fluid volume as the smooth-wall channel.

The ratio of pressure to total drag, figure \ref{fig:solidity}(\emph{b}), also shows a difference between the sparse and dense roughness regimes. The ratio appears to increase in the sparse regime in a log-linear manner, whereby the pressure drag increases with increasing solidity. Once the dense  regime is encountered ($\Lambda > 0.15$), the pressure drag plateaus at about 50\% of the total drag. The limit of $\Lambda\rightarrow\infty$, which is modelled as an offset smooth wall in the previous section, suggests that the ratio must decrease for extremely dense surfaces and tend to zero. This indicates that the crossover into the dense regime ($\Lambda\approx0.15$) is where the pressure drag is maximum.
The cube-mounted surfaces of \cite{Leonardi10} are also shown in figure \ref{fig:solidity}(\emph{b}), although here the cubes are in the fully rough regime where the pressure drag is dominant at around 90--95\% of the total drag. However, it can still be seen that the pressure drag reaches a maximum around $\Lambda\approx0.15$ and then slightly decreases as the solidity becomes progressively more dense. As described previously, the cubes of \cite{Leonardi10} become a smooth wall at $\Lambda=1$, which would correspond to zero pressure drag.

\setlength{\unitlength}{1cm}
\begin{figure}
\centering
 \captionsetup[subfigure]{labelformat=empty}

	\subfloat[]{
		\includegraphics[trim=0 5 0 0,clip=true]{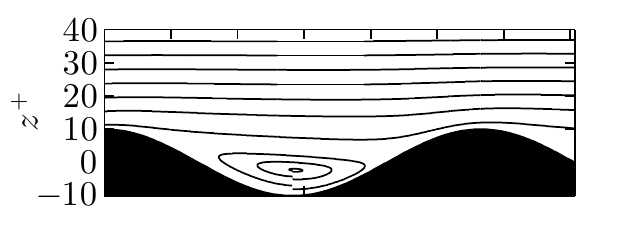}
	}
	\put(-3.35,1.50){\colorbox{white}{$\Lambda = 0.11$ (sparse)}} 
	\subfloat[]{
		\includegraphics[trim=0 5 0 0,clip=true]{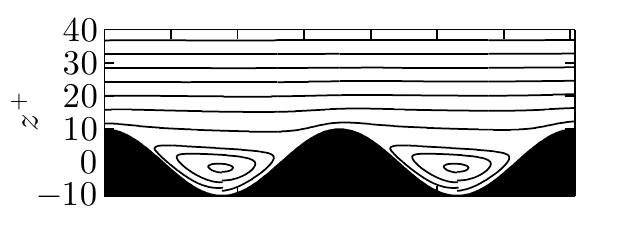}
	}
	\put(-3.25,1.50){\colorbox{white}{$\Lambda = 0.18$ (dense)}} 
	\put(-12.8,1.8){(\emph{a})}
	\put(-6.35,1.8){(\emph{b})}
	\\  \vspace{-2.8\baselineskip}
	\subfloat[]{
		\includegraphics[trim=0 0 0 0,clip=true]{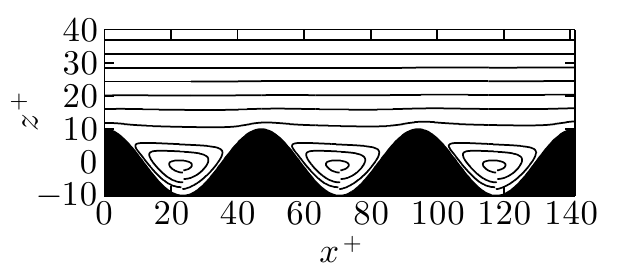}
	}
	\put(-3.25,2.05){\colorbox{white}{$\Lambda = 0.36$ (dense)}} 
	\subfloat[]{
		\includegraphics[trim=0 0 0 0,clip=true]{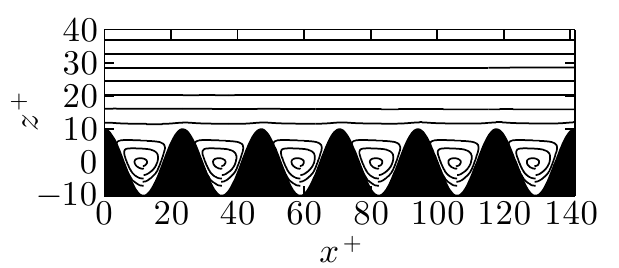}
	}
	\put(-3.25,2.05){\colorbox{white}{$\Lambda = 0.54$ (dense)}} 
	\put(-12.8,2.35){(\emph{c})}
	\put(-6.35,2.35){(\emph{d})}
	\vspace{-2.0\baselineskip}
\caption{Mean streamlines over
 (\emph{a}) $\lambda^+= 113\Rightarrow \Lambda=0.11$,
 (\emph{b}) $\lambda^+= 70.7\Rightarrow \Lambda=0.18$,
 (\emph{c}) $\lambda^+= 47.1\Rightarrow \Lambda=0.36$,
and
 (\emph{d}) $\lambda^+= 23.6\Rightarrow \Lambda=0.54$ roughness, in the streamwise--wall-normal plane.
Flow is in (\emph{a}) sparse and (\emph{b},\emph{c},\emph{d}) dense regimes.
}
\label{fig:streamlines}
\end{figure}
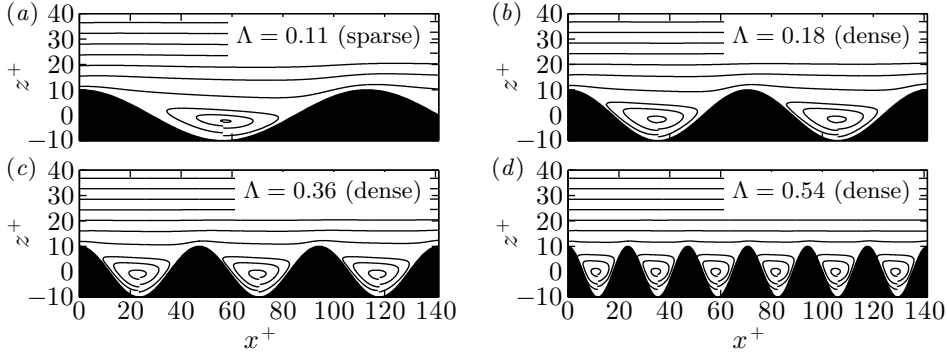

Conceptual models of the dense regime of roughness often describe stable vortices within the roughness elements,  with high-speed fluid skimming over the top of the roughness. The sparse regime, meanwhile, is described by a much smaller recirculation zone, with the separation point being closer to the reattachment point \citep{Oke88,Macdonald00}. In order to assess the veracity of these descriptions, the mean streamlines are shown in figure \ref{fig:streamlines} for both sparse and dense roughness. All four sets of streamlines show an almost identical flow pattern, with the recirculation region appearing similar in terms of the roughness wavelength. 
The area of flow recirculation, $A_R$ does scale with solidity according to $A_{R}/A_{T}\approx0.18\log(\Lambda)+0.9$, where $A_T$ is the total area occupied by fluid below the roughness crest, however there does not appear to be a distinct change in flow structure between the sparse (figure \ref{fig:streamlines}\emph{a}) and dense (figure \ref{fig:streamlines}\emph{b}--\emph{d}) regimes.
It is clear that these qualitative descriptions of roughness are not adequate on their own to indicate existence of the dense regime, or to explain why a slightly different flow pattern results in a reduction in the roughness function.

\setlength{\unitlength}{1cm}
\begin{figure}
\centering
 \captionsetup[subfigure]{labelformat=empty}

	\subfloat[]{
		\includegraphics{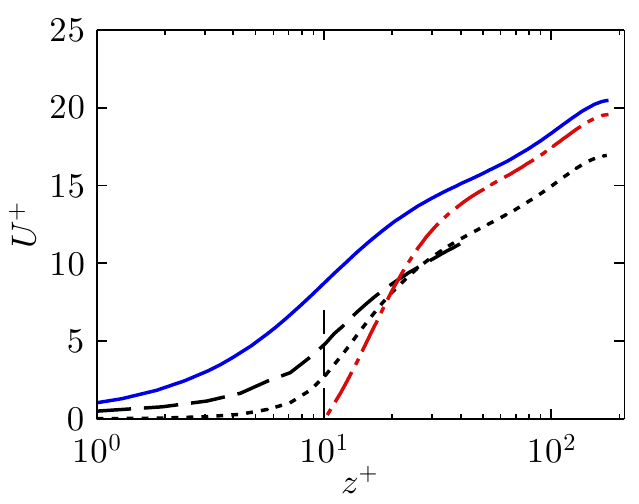}
	}
	\subfloat[]{
		\includegraphics{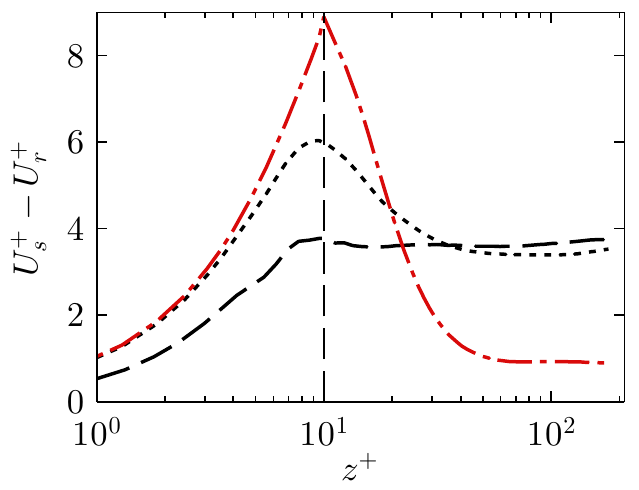}
	}
	\put(-13.45,4.7){(\emph{a})}
	\put(-10.3,2.0){\vector(1,-1){.8}}
	\put(-9.45,1.15){Increasing $\Lambda$}
	\put(-6.6,4.7){(\emph{b})}
	\put(-3.0,2.3){\vector(0,1){2.2}}
	\put(-2.85,4.3){Increasing $\Lambda$}
	\put(-1.5,2.75){$\Lambda$$=$$0.11$}
	\put(-1.5,2.1){$\Lambda$$=$$0.54$}
	\put(-1.5,1.45){$\Lambda$$\rightarrow$$\infty$}
	\vspace{-2.2\baselineskip}
\caption{(Colour online) (\emph{a}) Mean velocity profile and (\emph{b}) difference in smooth- and rough-wall velocity for minimal-span channels. Line styles:
\protect\raisebox{0.8ex}{\color{myblue}\linethickness{0.5mm}\line(1,0){0.6}}, smooth wall  ($\Lambda=0$);
\protect\raisebox{0.8ex}{\linethickness{0.5mm}\line(1,0){0.3}\hspace{0.15cm}\line(1,0){0.3}}, sparse roughness ($\Lambda=0.11$);
\protect\raisebox{0.8ex}{\linethickness{0.5mm}\line(1,0){0.1}\hspace{0.15cm}\line(1,0){0.1}\hspace{0.15cm}\line(1,0){0.1}}, dense roughness ($\Lambda=0.54$);
\protect\raisebox{0.8ex}{\color{myred}\linethickness{0.5mm}\line(1,0){0.08}\hspace{0.15cm}\line(1,0){0.2}\hspace{0.15cm}\line(1,0){0.08}}, offset smooth wall ($\Lambda\rightarrow\infty$).
Vertical dashed line shows roughness crest $k^+=10$.
Outer-layer profile in $(a)$ for sparse roughness ($\Lambda=0.11$) is omitted for clarity (see text).
}
	\label{fig:UxModSmooth}
\end{figure}

Figure \ref{fig:UxModSmooth}(\emph{a}) shows the mean velocity profile for the smooth wall ($\Lambda=0$), a sparse regime case ($\Lambda=0.11$), a dense regime case ($\Lambda=0.54$), and the offset smooth wall in which the wall is located at $z_w^+=10$  ($\Lambda\rightarrow\infty$). The sparse regime case has a different spanwise width ($L_y^+=113$) compared to the other cases ($L_y^+=141$), meaning the mean velocity profile above $z_c^+\approx45$ would be different (figure \ref{fig:velProfValid}). These differences are not relevant to the ensuing discussion on the effects of solidity, so for clarity the outer-layer profile for this sparse regime case is omitted above $z_c^+= 45$. The difference in smooth- and rough-wall velocity profiles (figure \ref{fig:UxModSmooth}\emph{b}) uses the same $L_y^+$ for both smooth and rough-wall flows which eliminates this dependence on $L_y^+$.

It can be seen in figure \ref{fig:UxModSmooth}(\emph{a}) that as the solidity increases, the near-wall streamwise velocity decreases. As such, the velocity difference between smooth- and rough-wall flows (figure \ref{fig:UxModSmooth}\emph{b}) increases  with increasing solidity close to the wall. The sparse roughness case (dashed line) has a constant velocity difference with the smooth wall from the crest of the roughness to the channel centre. However, the dense roughness case (dotted line) shows a rapid increase in velocity above the crest of the roughness, from $k^+<z^+\lesssim40$, which causes the velocity difference to decrease with $z^+$, until this difference is less than that of the sparse roughness case at around $z^+\approx40$. Above $z^+\gtrsim40$, the velocity over the  dense roughness is similar in shape to the smooth wall, so that the velocity difference is constant.
The end result is that the roughness functions of these two cases are nearly identical, with $\Delta U^+\approx3.6$ for the sparse case and $\Delta U^+\approx3.4$ for the dense case.
These matched values occur even though the dense roughness has nearly 5 times more roughness elements in the streamwise direction (compare the roughness depictions of figure \ref{fig:streamlines}\emph{a} and \ref{fig:streamlines}\emph{d}). Despite the substantial changes in the near-wall velocity, with the velocity being near zero within much of the roughness canopy for the dense roughness, the flow is able to reorganise in a way that leads to nearly identical roughness functions.
As mentioned earlier, this is reminiscent of the work by \cite{Antonia01} who had nominally matched $\Delta U^+$ for two markedly different surfaces. 
The limit of solidity is the offset smooth-wall flow, which has zero velocity at $z^+=k^+$ so the velocity difference is very large close to the wall. The velocity in this offset smooth-wall flow rapidly increases in the near-wall region until $z^+\approx 40$, at which point it remains similar to the smooth-wall velocity and so the velocity difference is constant. Even in this limit of solidity, the velocity difference is constant above $z^+\gtrsim40$ which is below the critical wall-normal location $z_c^+=0.4L_y^+\approx56$ of healthy turbulence that is captured by the minimal-span channel.

\setlength{\unitlength}{1cm}
\begin{figure}
\centering
 \captionsetup[subfigure]{labelformat=empty}

	\subfloat[]{
		\includegraphics[width=0.49\textwidth]{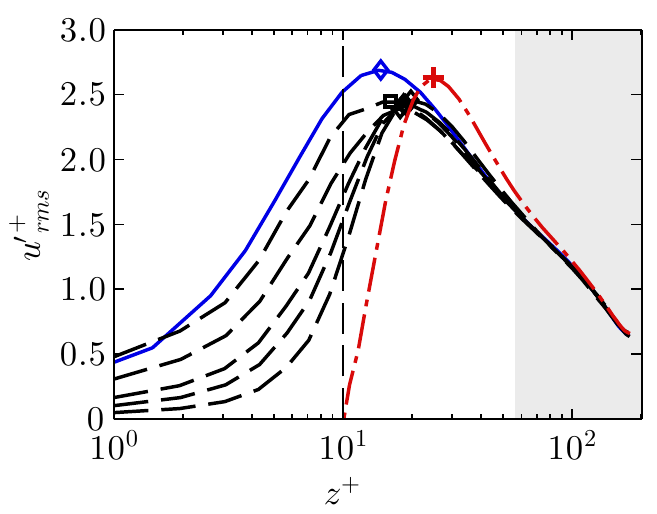}
	}
	\subfloat[]{
		\includegraphics[width=0.5\textwidth]{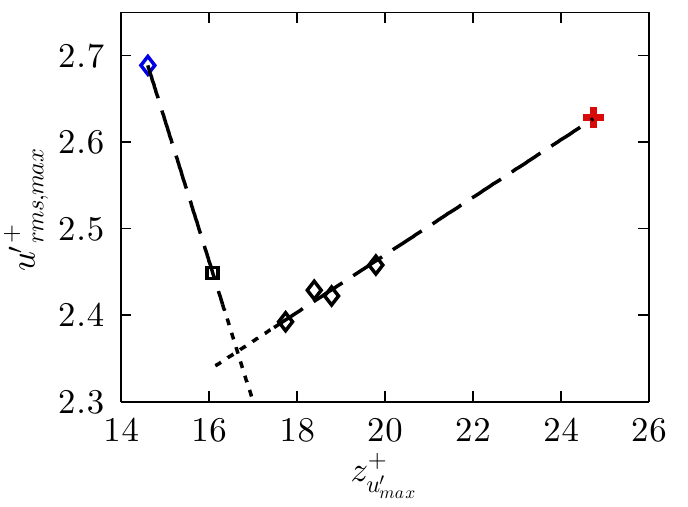}
	}
	\put(-13.65,4.85){(\emph{a})}
	\put(-11.0,2.6){\vector(1,-1){1}}
	\put(-9.9,1.5){Increasing $\Lambda$}
	\put(-8.36,0.97){\line(0,1){0.5}}
	\put(-8.33,1.1){$z_c^+$=$0.4L_y^+$}
	\put(-6.7,4.85){(\emph{b})}
	\put(-5.40,4.35){$\Lambda$$=$$0$}
	\put(-4.7,2.38){$\Lambda$$=$$0.11$}
	\put(-3.85,1.5){\vector(3,2){1.1}}
	\put(-3.4,1.5){Increasing $\Lambda$$\geq$$0.18$}
	\put(-1.45,3.85){$\Lambda$$\rightarrow$$\infty$}
	\vspace{-3.0\baselineskip}
	\\
	\subfloat[]{
		\includegraphics[width=0.49\textwidth]{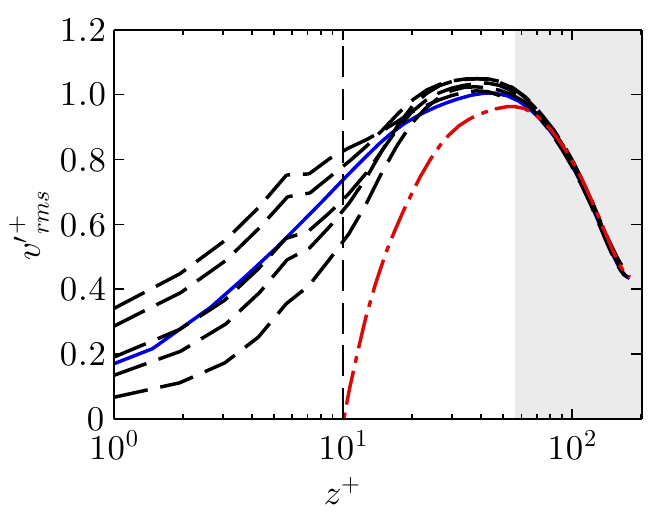}
	}
	\subfloat[]{
		\includegraphics[width=0.49\textwidth]{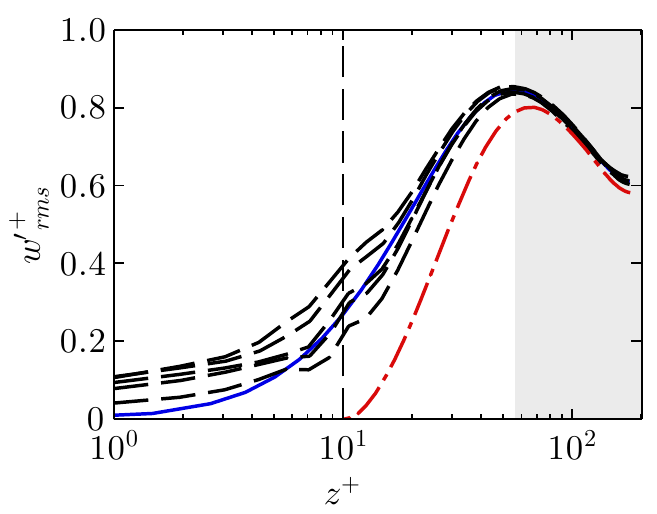}
	}
	\put(-13.55,4.85){(\emph{c})}
	\put(-11.2,3.1){\vector(1,-1){1}}
	\put(-10.05,2.0){Increasing $\Lambda$}
	\put(-8.36,0.97){\line(0,1){0.5}}
	\put(-8.33,1.1){$z_c^+$=$0.4L_y^+$}
	\put(-6.7,4.85){(\emph{d})}
	\put(-3.5,2.5){\vector(1,-1){0.75}}
	\put(-2.7,1.7){Increasing $\Lambda$}
	\put(-1.53,0.97){\line(0,1){.5}}
	\put(-1.50,1.1){$z_c^+$=$0.4L_y^+$}
	\vspace{-2.2\baselineskip}	
\caption{(Colour online) Root-mean-square velocity fluctuations in the (\emph{a}) streamwise, (\emph{c}) spanwise, and (\emph{d}) wall-normal directions for minimal-span channels. Line styles: 
\protect\raisebox{0.8ex}{\color{myblue}\linethickness{0.5mm}\line(1,0){0.6}}, smooth wall  ($\Lambda=0$);
\protect\raisebox{0.8ex}{\linethickness{0.5mm}\line(1,0){0.3}\hspace{0.15cm}\line(1,0){0.3}}, rough wall with solidity $0.11\leq\Lambda\leq0.54$, refer to table \ref{tab:sims}; 
\protect\raisebox{0.8ex}{\color{myred}\linethickness{0.5mm}\line(1,0){0.08}\hspace{0.15cm}\line(1,0){0.2}\hspace{0.15cm}\line(1,0){0.08}}, offset smooth wall ($\Lambda\rightarrow\infty$).
Vertical dashed line shows roughness crest, $k^+=10$.
Shaded region shows $z_c^+=0.4L_y^+$ (figure \ref{fig:velProfValid}).
(\emph{b}) Maximum value of $u'^+_{rms}$ against its corresponding wall-normal location. Symbols are:
$\color{myblue} \boldsymbol\diamondsuit$, smooth wall ($\Lambda = 0$);
$\boldsymbol\square$, sparse regime ($\Lambda = 0.11$);
$\boldsymbol\diamondsuit$, dense regime ($0.18\leq\Lambda\leq0.54$);
\protect\raisebox{0.4ex}{\protect\scalebox{1.0}{$\color{myred}\boldsymbol{\pmb{+}}$}}, offset smooth wall ($\Lambda\rightarrow\infty$).
Dashed line comes from simple linear regression. 
}
	\label{fig:velFluc}
\end{figure}

The root-mean-square velocity fluctuations in the streamwise, spanwise, and wall-normal directions are shown in figure \ref{fig:velFluc}(\emph{a,c,d}) for all the minimal-span channel data. Here, it can be seen that as the solidity increases, the velocity fluctuations close to the wall decrease and tend towards the offset smooth-wall data. The sparser roughness cases exhibit  spanwise  and wall-normal velocity fluctuations (figure \ref{fig:velFluc}\emph{c},\emph{d}) that are enhanced compared to that of the smooth wall. As the solidity increases and the roughness enters the dense regime, the spanwise and wall-normal velocity fluctuations are damped compared to the smooth wall. It appears that the sparse roughness increases the spanwise and wall-normal velocity fluctuations within the roughness canopy, yet as the roughness becomes increasingly dense, these velocity fluctuations are increasingly damped and are tending towards the limiting case of $\Lambda\rightarrow\infty$. The streamwise velocity fluctuations are damped compared to the smooth wall for both sparse and dense regimes.
The roughness function has been observed by \cite{Orlandi06jfm} to scale monotonically with the value of $w'^+_{rms}$ at the crest of the roughness, at least for two-dimensional bar roughness. However, for the present experiments this behaviour is not observed, as $w'^+_{rms}$ at the roughness crest can be seen to decrease monotonically with $\Lambda$  (figure \ref{fig:velFluc}\emph{d}), yet $\Delta U^+$ is not monotonic with $\Lambda$ (figure \ref{fig:solidity}\emph{a}). This is the same result as \cite{Leonardi10}, where they suggested that the three-dimensional nature of the roughness that they studied may be why the roughness function was not monotonic with respect to $w'^+_{rms}$.

\setlength{\unitlength}{1cm}
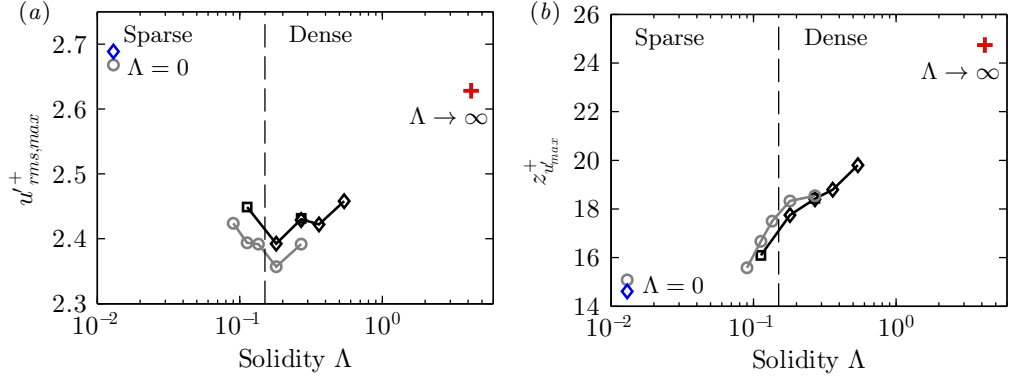
\begin{figure}
\centering
 \captionsetup[subfigure]{labelformat=empty}

	\subfloat[]{
		\includegraphics[width=0.49\textwidth]{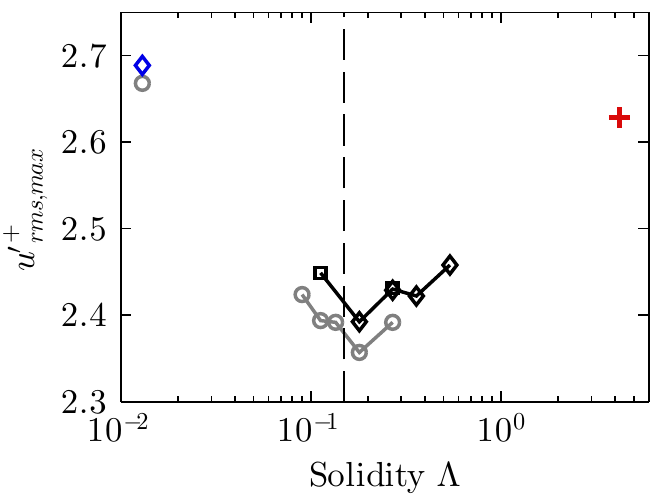}
	}
	\subfloat[]{
		\includegraphics[width=0.49\textwidth]{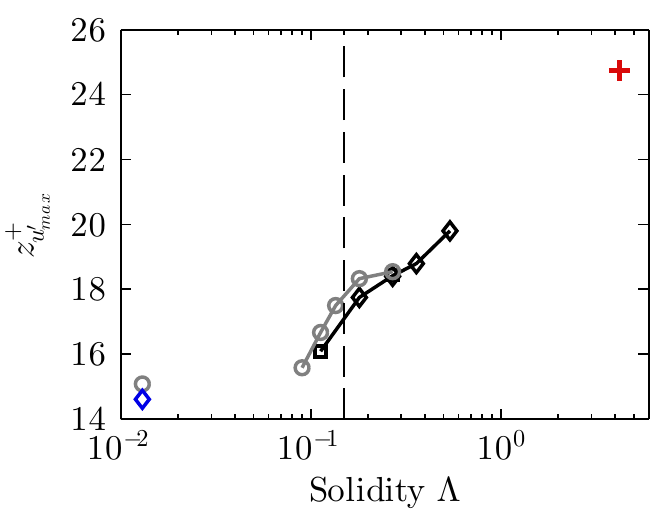}
	}
	\put(-13.4,4.8){(\emph{a})}
	\put(-12.0,4.5){Sparse}
	\put(-9.8,4.5){Dense}
	\put(-11.95,4.05){$\Lambda=0$}
	\put(-8.2,3.4){$\Lambda\rightarrow\infty$}
	\put(-6.6,4.8){(\emph{b})}
	\put(-5.2,4.45){Sparse}
	\put(-2.95,4.45){Dense}
	\put(-5.1,1.15){$\Lambda=0$}
	\put(-1.4,4.0){$\Lambda\rightarrow\infty$}
	\vspace{-2.2\baselineskip}	
\caption{(\emph{a}) Maximum value of streamwise turbulence intensity, $u'^+_{rms,max}$ and (\emph{b}) its corresponding wall-normal location, against solidity $\Lambda$. Symbols are same as figure \ref{fig:solidity}. Rough-wall pipe data (grey circles) only shown for $\Lambda\ge0.09$ (see text).
}
	\label{fig:FlucSol}
\end{figure}

 Figure \ref{fig:velFluc}(\emph{b}) shows the maximum value of the streamwise velocity fluctuations, $u'^+_{rms,max}$ against the wall-normal position where this maximum occurs. The smooth-wall flow ($\Lambda = 0$) show the maximum value occurring at a wall-normal position of $z^+_{u'_{max}}\approx15$, which corresponds to the location of the near-wall cycle.  We see that the sparse regime ($\Lambda=0.11$) causes a reduction in the peak value of the streamwise velocity fluctuation compared to the smooth-wall flow, as well as moving the peak location away from the wall. In the dense regime, increasing the solidity over the range $0.18\leq \Lambda\leq0.54$ causes the maximum streamwise velocity fluctuation to move increasingly further from the wall, as well as increasing the maximum value of streamwise velocity fluctuations.
This is more clearly presented in figure \ref{fig:FlucSol}(\emph{a}), where the peak streamwise velocity fluctuations are shown as a function of solidity. Both the pipe (grey) and minimal-span channel (black) data indicate that increasing solidity in the sparse regime reduces the peak streamwise turbulent energy, while increasing solidity in the dense regime leads to an increase in energy. This trend in the dense regime appears to be tending towards the offset smooth-wall flow ($\Lambda\rightarrow\infty$), where the peak energy is similar to that of the smooth wall ($\Lambda=0$).
Figure \ref{fig:FlucSol}(\emph{b}) shows the wall-normal location of the peak streamwise turbulent energy against solidity. Following figure \ref{fig:velFluc}(\emph{b}), the location of the peak is moving increasingly outwards, irrespective of the sparse or dense roughness regimes. The pipe flow data only shows rough walls with a solidity greater than 0.09 ($\lambda^+\le141$), as sparser roughness results in the peak streamwise energy being located below the roughness crest (figure 18 of \citealt{Chan15}).

\setlength{\unitlength}{1cm}
\begin{figure}
\centering
 \captionsetup[subfigure]{labelformat=empty}

	\subfloat[]{
		\includegraphics[width=0.49\textwidth]{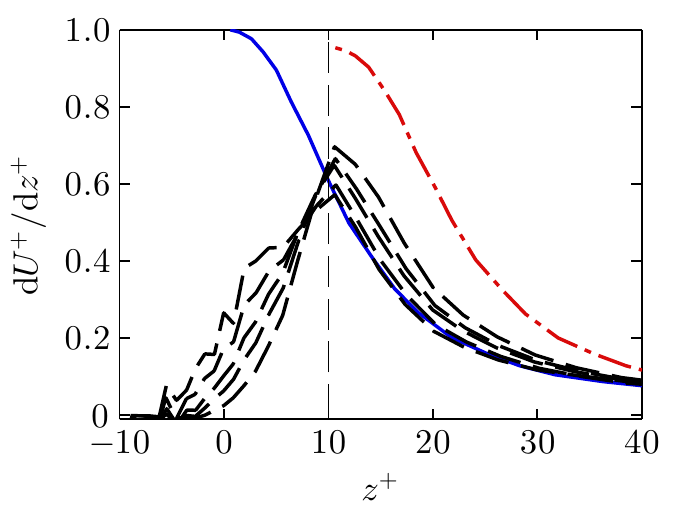}
	}
	\subfloat[]{
		\includegraphics[width=0.49\textwidth]{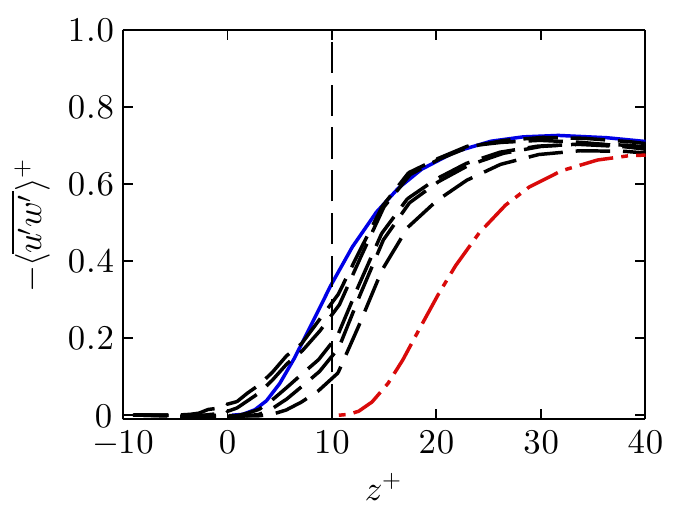}
	}
	\put(-13.55,4.55){(\emph{a})}
	\put(-10.25,2.5){\vector(1,2){.5}}
	\put(-11.4,2.3){\vector(1,-1){.75}}
	\put(-9.7,3.5){Increasing $\Lambda$}
	\put(-12.0,4.3){$\Lambda = 0$}
	\put(-9.85,4.3){$\Lambda \rightarrow\infty$}
	\put(-6.7,4.55){(\emph{b})}
	\put(-3.4,3){\vector(1,-2){.5}}
	\put(-2.85,1.8){Increasing $\Lambda$}
	\vspace{-3.0\baselineskip}
	\\
	\subfloat[]{
		\includegraphics[width=0.49\textwidth]{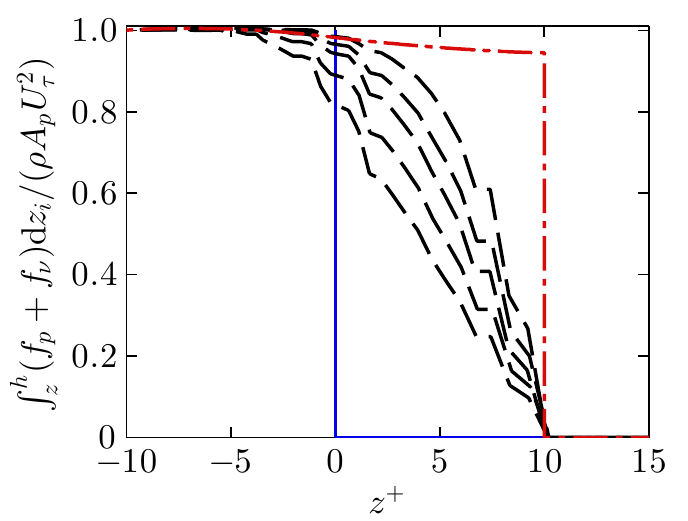}
	}
	\subfloat[]{
		\includegraphics[width=0.475\textwidth]{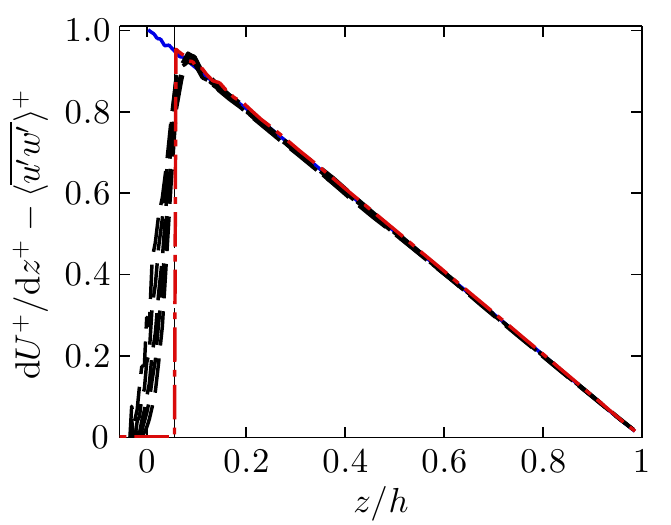}
	}
	\put(-13.3,4.65){(\emph{c})}
	\put(-9.7,2.8){\vector(1,2){0.7}}
	\put(-8.95,4.15){Increasing $\Lambda$}
	\put(-6.5,4.65){(\emph{d})}
	\put(-5.2,2.5){\vector(2,-1){0.6}}
	\put(-4.55,2.1){Increasing $\Lambda$}
	\vspace{-2.2\baselineskip}
\caption{
(Colour online)
(\emph{a}) Viscous stress, 
(\emph{b}) Reynolds stress, 
(\emph{c}) integrated total drag per unit wall-normal distance,
and
(\emph{d}) total stress (viscous (\emph{a}) + Reynolds (\emph{b}) stress) profiles.
The sum of (\emph{c}) and (\emph{d}) is the integral of the streamwise momentum balance, which will be 1 at the roughness trough.
Line styles are same as figure \ref{fig:velFluc}.
}
	\label{fig:stress}
\end{figure}
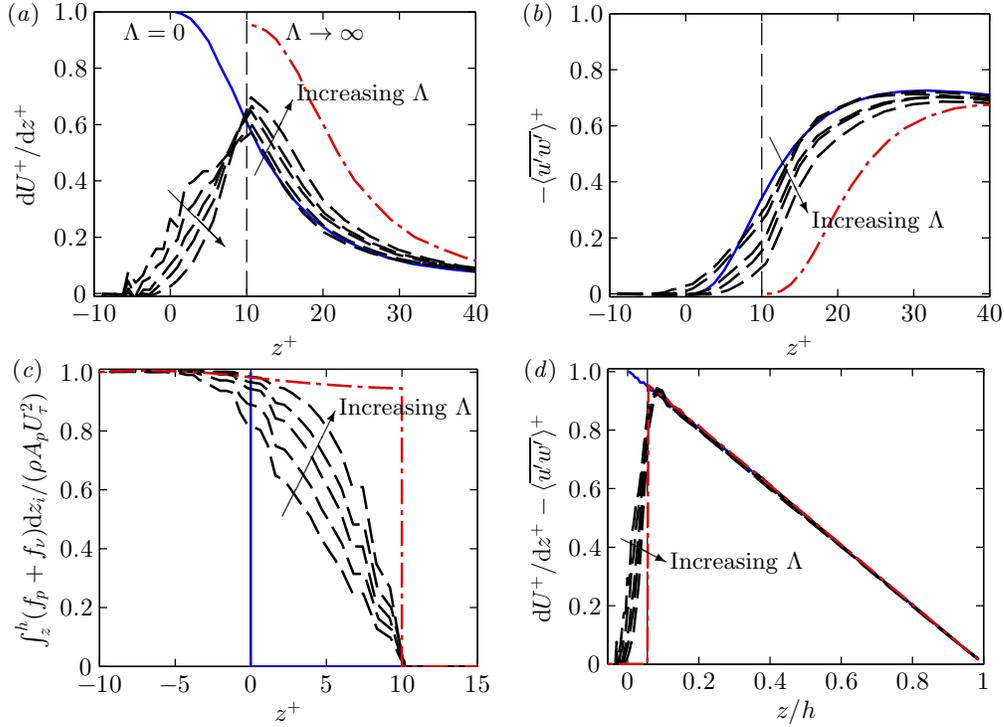

The viscous and Reynolds stress terms are shown in figure \ref{fig:stress}.
Understanding these stress terms, and how they vary with roughness, is important for the characterisation of turbulent flows as well as eddy-viscosity models.
Increasing the roughness solidity results in these stresses tending towards the offset smooth-wall data. This results in the dense roughness data having decreased viscous stress (figure \ref{fig:stress}\emph{a}) below the crest of the roughness, and enhanced viscous stress above the crest, relative to the smooth-wall flow. 
Above $z^+\gtrsim40$ there is an apparent collapse for smooth-wall and rough-wall flows, although this will not be preserved in the limit of $\Lambda\rightarrow\infty$.  The Reynolds stress (figure \ref{fig:stress}\emph{b}) is seen to decrease as the solidity increases, as the roughness dampens out all three velocity fluctuation components (figure \ref{fig:velFluc}), again tending towards the limiting case.
The sum of the pressure and viscous drag per unit wall-normal distance, $f_p$ and $f_\nu$, can be integrated from $z$ to the channel centre to indicate what proportion of the total drag has been exerted above $z$. When normalised on $\rho A_p U_\tau^2$ (here $A_p$ is the wall-plan area), the integral $\int_z^h(f_p+f_\nu)\id z_i/(\rho U_\tau^2 A_p)$, where $z_i$ is a dummy variable, must be 1 at the roughness trough and 0 at the crest. This is plotted in figure \ref{fig:stress}(\emph{c}) and it can be seen that as solidity increases, the majority of the drag is exerted closer to the crest of the roughness. In the limit, nearly all ($1-k/h$) of the drag is applied at the crest in the form of viscous drag, with the small remainder ($k/h$) being distributed as pressure drag due to the driving pressure gradient applied to the channel (\S\ref{sect:dense}). The total fluid stress (sum of viscous (\emph{a}) and Reynolds (\emph{b}) stresses) is shown in figure \ref{fig:stress}(\emph{d}) as a function of $z/h$, and can be seen to linearly vary to be zero at the channel centre ($z/h=1$). Within the roughness canopy, the total fluid stress reduces to zero as the drag force (\emph{c}) is now being exerted across $-k < z < k$, rather than at a single point as in a smooth wall. The sum of the profiles of the wall drag per unit plan area (figure \ref{fig:stress}\emph{c}) and the total fluid stress (figure \ref{fig:stress}\emph{d}) is  the integral of the streamwise momentum balance, which will collapse with the sum of the limiting case ($\Lambda\rightarrow\infty$, red dash-dotted line).


\subsection{Mean momentum balance}
\label{sect:mmb}
\cite{Fukagata02} demonstrated that analysing the terms in the mean momentum balance gave insights into the dynamics of fluid flow, and showed it could be used to analyse the contributions of different Reynolds number effects on the skin-friction drag. Following along these lines, the integrated mean momentum balance for turbulent fluid flow is given below,
normalised on $\nu$ and $U_\tau$,
\begin{subequations}
\label{eqn:mmb}
\begin{align}
\tau_{uw,s}^+ + \frac{\dd U_s^+}{\dd z^+} &= \frac{h^+ -z^+}{h^+} \label{eqn:mmbS} \\
\tau_{uw,r}^+ + \frac{\dd U_r^+}{\dd z^+} &= \frac{h'^+ -z^+}{h'^+},\label{eqn:mmbR}
\end{align}
\end{subequations}
where subscript $s$ is used to denote smooth-wall flow, and subscript $r$ for rough-wall flow in which
the rough-wall equation (\ref{eqn:mmbR}) is only valid above the roughness crest, $z^+>k^+$.
Here, $\tau_{uw}^+=-\langle{\overline{u'w'}}\rangle^+$ is the Reynolds shear stress and $h'$ is the rough-wall channel half-height. Essentially, the Reynolds and viscous stress in the fluid gives rise to the linear trend seen in the total stress, which is unity at the wall in smooth-wall flows and zero at the channel centre. If the friction Reynolds number is matched between smooth- and rough-wall flows, then $h'^+=h^+$ and the right hand side of  (\ref{eqn:mmbS}) and (\ref{eqn:mmbR}) is therefore the same. This implies that the total stress profile collapses for the two flows.

\setlength{\unitlength}{1cm}
\begin{figure}
\centering
 \captionsetup[subfigure]{labelformat=empty}

	\subfloat[]{
		\includegraphics[width=0.49\textwidth]{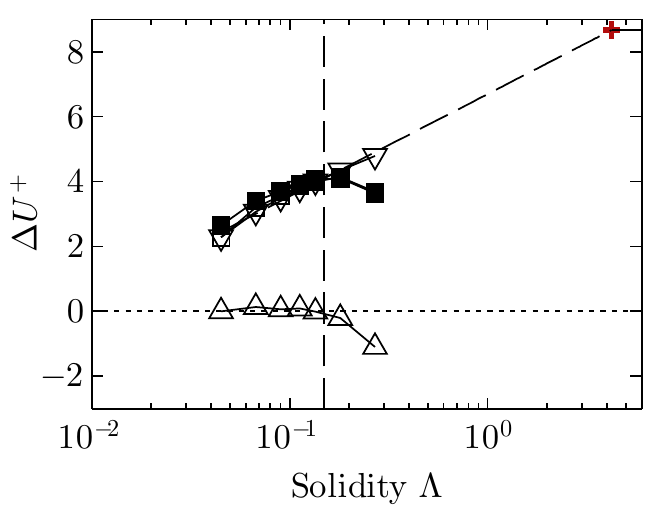}
	}
	\subfloat[]{
		\includegraphics[width=0.49\textwidth]{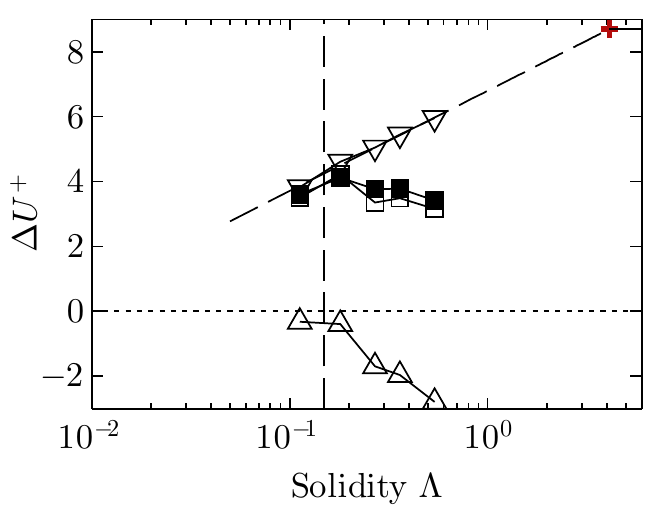}
	}
	\vspace{-2.0\baselineskip}
	\put(-13.55,4.85){(\emph{a})}
	\put(-6.75,4.85){(\emph{b})}
	\put(-12.4,4.5){Full-span pipe}
	\put(-5.6,4.5){Min.-span channel}
	\put(-9.6,3.4){$T_1$}
	\put(-9.6,1.6){$T_2$}
	\put(-12.4,1.12){Sparse}
	\put(-10.2,1.12){Dense}
	\put(-2.2,3.7){$T_1$}
	\put(-2.2,1.2){$T_2$}
	\put(-5.6,1.12){Sparse}
	\put(-3.45,1.12){Dense}
\caption{Contributions $T_1$ and $T_2$ to the roughness function $\Delta U^+$, equation (\ref{eqn:DU_decomp}), for (\emph{a}) pipe flow and (\emph{b}) minimal-span channel. Line styles:
\protect\raisebox{0.4ex}{\protect\scalebox{1.0}{$\bigtriangledown$}}, $T_1$;
$\bigtriangleup$, $T_2$, corrected with the observations made in the appendix;
$\square$, $T_1 + T_2$;
$\blacksquare$, $\Delta U^+$ computed from the mean velocity profile;
\protect\raisebox{0.4ex}{\protect\scalebox{1.0}{\color{myred}$\boldsymbol{\pmb{+}}$}}, $T_1$ for limit of $\Lambda\rightarrow\infty$.
Dashed line shows $T_1\approx 1.36\log(\Lambda)+6.7$.
Vertical dashed line demarcates sparse and dense regimes, $\Lambda = 0.15$.
}
	\label{fig:DU_decomp}
\end{figure}

Integrating (\ref{eqn:mmb}) from the roughness crest to the channel centre, $k^+<z^+\le h^+$ and equating the two equations,
\begin{equation}
\label{eqn:mmb_eq}
\int_{k^+}^{h^+}\tau_{uw,r}^+ \id z^++\left(U_{h r}^+-U_{k r}^+\right) = 
\int_{k^+}^{h^+}\tau_{uw,s}^+ \id z^++\left(U_{h s}^+-U_{k s}^+\right),
\end{equation}
where $U_{k}=U(z=k)$ and $U_{h}=U(z=h)$ is the streamwise velocity at the height of the roughness crest and at the channel centre, respectively. We can now rearrange for the roughness function $\Delta U^+$,
\begin{equation}
\label{eqn:DU_decomp}
\Delta U^+ = U_{h s}^+ - U_{h r}^+ = \underbrace{U_{ks}^+-U_{kr}^+\vphantom{\int_k^h}}_{T_1}
\hspace{1em} + \hspace{1em}
\underbrace{\int_{k^+}^{h^+}\left(\tau_{uw,r}^+-\tau_{uw,s}^+\right)\id z^+}_{T_2}.
\end{equation}
The terms $T_1$ and $T_2$ are similar to the second and third terms, respectively, of \cite{Mayoral11jfm}, used in the context of drag reduction. 
The first term, $T_1$, represents the difference in streamwise velocity between smooth- and rough-wall flows, at the height of the roughness crest $z^+=k^+$. In general, the rough-wall velocity, $U_{kR}^+$ decreases with solidity (figure \ref{fig:UxModSmooth}\emph{a}). Hence, since $U_{kS}^+$ is fixed, the difference between the two, $T_1$, increases with solidity. The second term, $T_2$, is the change in Reynolds stress of the flow over a rough wall compared to over a smooth wall (the difference between the areas under the solid and dashed curves in figure \ref{fig:stress}\emph{b}).
The same analysis has been performed in the appendix assuming a different $Re_\tau$ between the smooth- and rough-wall flows. The flow quantities $\tau_{uw}$ and $U$ of the smooth- and rough-wall simulations are non-dimensionalised on their respective viscosity and friction velocities and the only difference is that (\ref{eqn:DU_decomp}) has an additional error term that is proportional to the difference in friction Reynolds numbers. For the current data, this is less than 2\% so is negligible.

The contributions of the $T_1$ and $T_2$ terms are shown in figure \ref{fig:DU_decomp}. The sum of $T_1$ and $T_2$  (open squares) is seen, for both pipe and minimal-span channel domains, to agree well with the roughness function $\Delta U^+$ computed from the mean velocity profiles of smooth and rough-wall flows (solid squares). 
The $T_2$ term (integrated difference in Reynolds shear stress) is corrected due to slight differences in the friction Reynolds number, which involves ignoring the outer-layer contribution to $T_2$. This is because Townsend's outer-layer similarity hypothesis predicts that $\tau_{uw}$ should collapse in the outer-layer (so the difference is zero), however small variations in $Re_\tau$ means this is not necessarily the case. For further details, refer to the appendix.   Note that even without the correction employed, the same trends involving the $T_2$ term are still observed (figure \ref{fig:T2_int} in the appendix). 
The $T_1$ term increases monotonically with increasing solidity. The primary cause of the reduction in roughness function in the dense roughness regime is therefore due to the reduction in Reynolds shear stress over the densely packed rough walls, $T_2$. This is clear from both the (full-span) pipe and minimal-span channel data, which both show the $T_2$ term reducing with increasing solidity for $\Lambda\gtrsim0.15$. In the sparse regime, $T_2$ is zero, indicating little change in the Reynolds shear stress between smooth-wall and sparse roughness flow.

The contributions of $T_1$ and $T_2$ can easily be seen when considering the difference in smooth- and rough-wall streamwise velocity as a function of $z^+$, as in figure \ref{fig:UxModSmooth}(\emph{b}). The difference in smooth- and rough-wall velocity remains constant above the crest of the roughness for the sparse roughness case (dashed line of figure \ref{fig:UxModSmooth}\emph{b}) so that $T_1 = U_{ks}^+-U_{kr}^+=\Delta U^+$ and there is a negligible $T_2$ contribution. 
 However, the dense regime has a value of $T_1$ that is greater than the roughness function. Hence $U_s^+ - U_r^+$ (dotted line of figure \ref{fig:UxModSmooth}\emph{b}) decreases  with increasing $z^+$ in the near wall region $k^+ < z^+ < 30$. This variation of $U_s^+-U_r^+$ with $z^+$ is balanced with a reduced Reynolds stress for the rough wall in the region $k^+<z^+<30$ (figure \ref{fig:stress}\emph{b}).

We have assumed that the offset smooth-wall flow represents the limit of $\Lambda\rightarrow\infty$, where the wall is located at the roughness crest $z_w^+=k^+$. Therefore there is zero velocity at the crest, $U_{kr}^+=0$, and the limiting value of $T_1$ is $T_1\rightarrow U_{ks}^+$, which has a value of 8.8 for the current flow. The $T_1$ term in figure \ref{fig:DU_decomp} appears to increase in a log-linear manner with $T_1\approx1.36\log(\Lambda)+6.7$. This approximation is seen to reach the limiting value of $T_1\approx8.8$  when the solidity is 4.2, corresponding to a sinusoidal wavelength of $\lambda^+\approx3$ when $k^+=10$ at $Re_\tau=180$. 
While this is a simple approximation, it supports the view that roughness with a length scale on the order of the viscous length scale ($\lambda^+\approx 1$) does not have an effect on the flow, apart from the offset in the wall location.

\setlength{\unitlength}{1cm}
\begin{figure}
\centering
 \captionsetup[subfigure]{labelformat=empty}

	\subfloat[]{
		\includegraphics[trim=0 8 0 0,clip=true,width=0.49\textwidth]{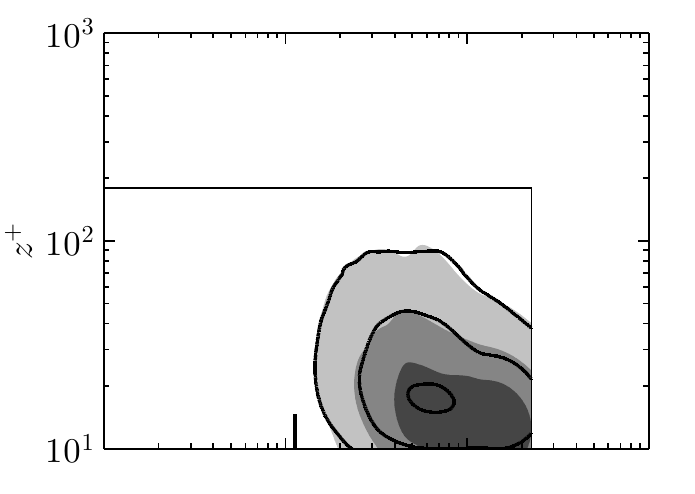}
	}
	\put(-4.46,0.25){\rotatebox{90}{\colorbox{white}{$\Lambda = 0.11$ (sparse)}}} 
	\subfloat[]{
		\includegraphics[trim=0 8 0 0,clip=true,width=0.49\textwidth]{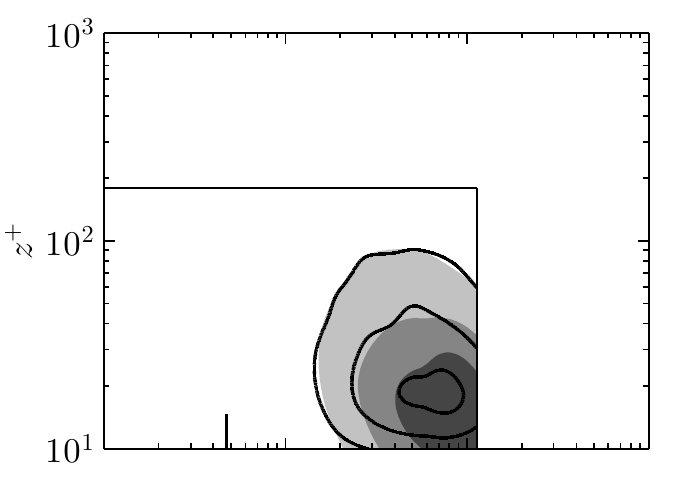}
	}
	\put(-5.12,0.25){\rotatebox{90}{\colorbox{white}{$\Lambda = 0.27$ (dense)}}} 
	\put(-13.65,4.1){(\emph{a})}
	\put(-6.85,4.1){(\emph{b})}
	\\  \vspace{-2.55\baselineskip}
	\subfloat[]{
		\includegraphics[trim=0 0 0 0,clip=true,width=0.49\textwidth]{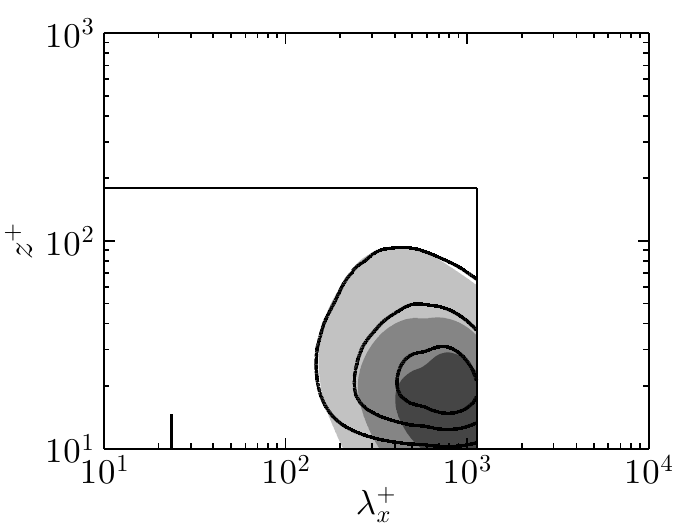}
	}
	\put(-5.66,0.85){\rotatebox{90}{\colorbox{white}{$\Lambda = 0.54$ (dense)}}} 
	\subfloat[]{
		\includegraphics[trim=0 0 0 0,clip=true,width=0.49\textwidth]{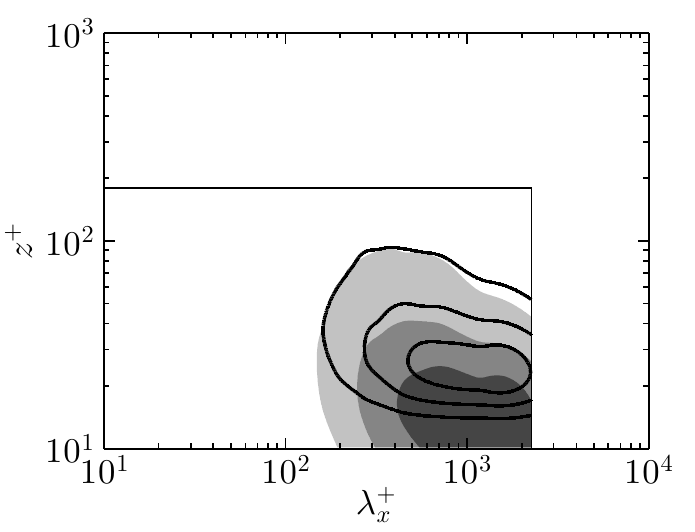}
	}
	\put(-5.66,0.85){\rotatebox{90}{\colorbox{white}{$\Lambda \rightarrow\infty$}}}
	\put(-13.65,4.65){(\emph{c})}
	\put(-6.85,4.65){(\emph{d})}
	\vspace{-2.0\baselineskip}
\caption{Streamwise pre-multiplied one-dimensional energy spectra of streamwise velocity for smooth (shaded) and rough walls (line).
Roughness wavelength 
 (\emph{a}) $\lambda^+=113$ (sparse),
 (\emph{b}) $\lambda^+=47.1$ (dense),
 (\emph{c}) $\lambda^+=23.6$ (dense),
and (\emph{d}) $\lambda^+\rightarrow0$, limit of $\Lambda\rightarrow\infty$ (offset smooth-wall located at $z_w^+=k^+$).
Contour levels are evenly spaced, corresponding to $k_x E_{uu}^+ = (0.6, 1.2, 1.8)$. Small vertical line at $z^+=10$ shows the roughness wavelength.
}
\label{fig:EuuContour}
\end{figure}
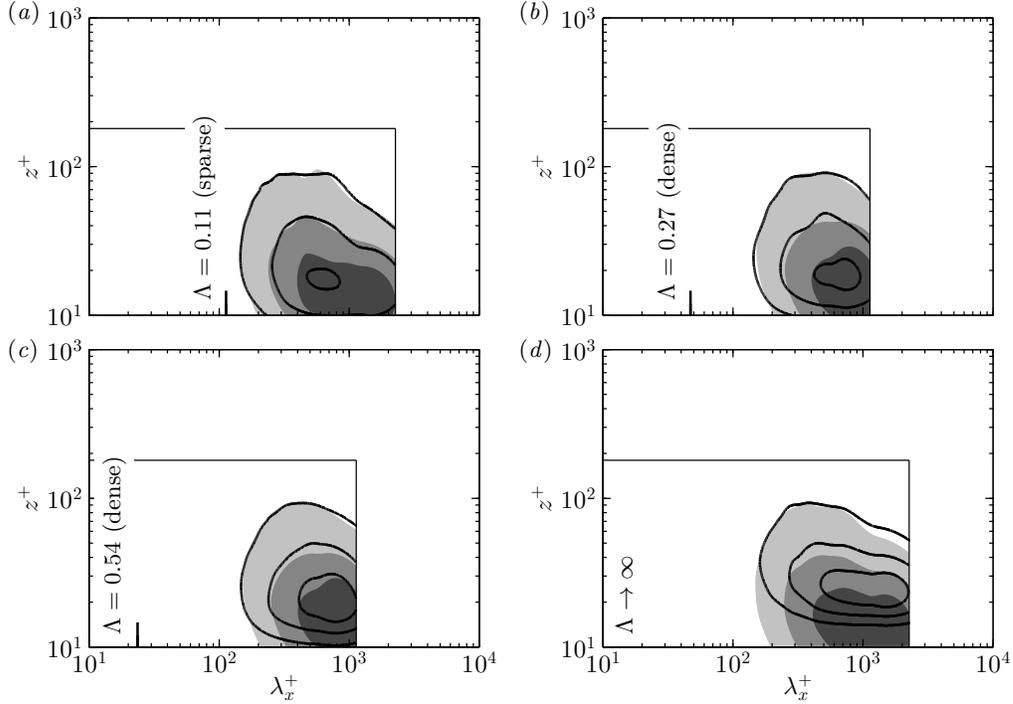

In order to see what flow changes are occurring in the dense regime, the one-dimensional pre-multiplied energy spectra of streamwise velocity is shown for smooth- and rough-wall flows in figure \ref{fig:EuuContour} by the filled and line contours, respectively. The sparse regime (figure \ref{fig:EuuContour}\emph{a}) is seen to have reasonable agreement between the smooth- and rough-wall flows, although the peak energy is smaller over the rough wall. This was previously observed in the peak streamwise turbulence intensity, figure \ref{fig:FlucSol}(\emph{a}), which is the integral of the energy spectra in figure \ref{fig:EuuContour}. As the solidity increases, the near-wall behaviour noticeably changes for the dense regime cases. The location of the peak rough-wall streamwise energy appears to move away from the wall as the solidity increases, as well as increasing in wavelength.
 There is therefore a loss of energy in the region previously occupied by the near-wall cycle ($z^+\approx15$) so that the densest roughness case (figure \ref{fig:EuuContour}\emph{c}) has significantly reduced energy immediately above the wall ($k^+< z^+\lesssim20$), compared to the smooth wall. In the limit of $\Lambda\rightarrow\infty$, the resultant flow is simply that of a smooth wall at $Re_\tau=170$ but offset so the wall is at $z_w^+=k^+$. As such, the near-wall cycle would have moved up by $k^+$ wall units but would otherwise be identical to the smooth-wall flow.
Figure \ref{fig:EuuContour}(\emph{d}) shows the offset smooth-wall flow, and it is clear that the near-wall cycle has been moved up by $k^+=10$ wall-units.  The location of the peak energy (not shown) agrees with the location of the peak streamwise turbulence intensity, shown in figure \ref{fig:FlucSol}(\emph{b}).

\setlength{\unitlength}{1cm}
\begin{figure}
\centering
 \captionsetup[subfigure]{labelformat=empty}

	\subfloat[]{
		\includegraphics[trim=0 0 0 0,clip=true,scale=1.0]{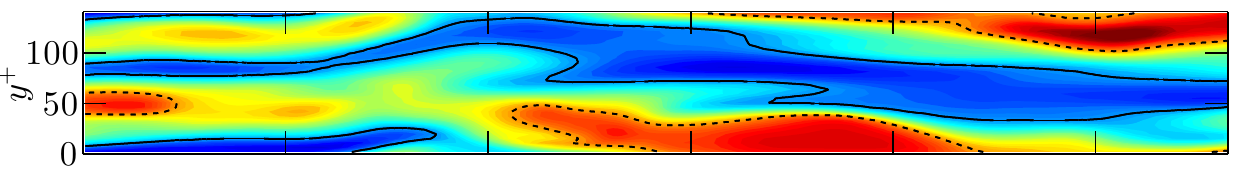}
	}
	\put(-3.2,1.1){\colorbox{white}{Smooth, $z^+\approx 14.3$}}
	\\ \vspace{-3.0\baselineskip}
	\subfloat[]{
		\includegraphics[trim=0 0 0 0,clip=true,scale=1.0]{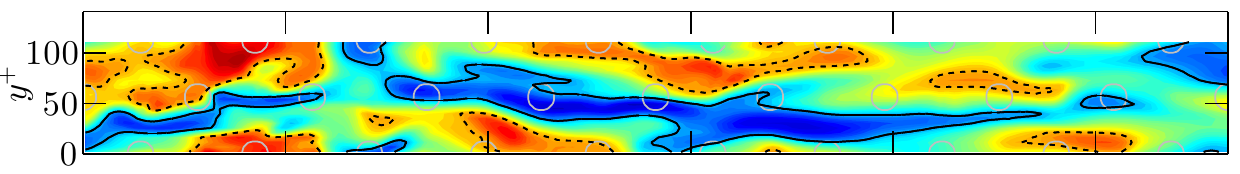}
	}
	\put(-4.6,1.1){\colorbox{white}{$\lambda^+ = 113$ (sparse), $z^+\approx 16.1$}}
	\\ \vspace{-3.0\baselineskip}
	\subfloat[]{
		\includegraphics[trim=0 0 0 0,clip=true,scale=1.0]{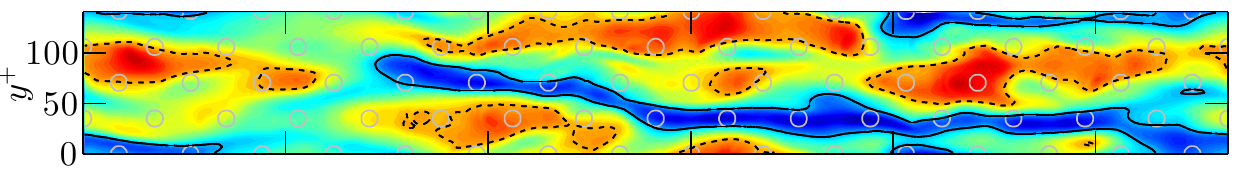}
	}
	\put(-4.6,1.12){\colorbox{white}{$\lambda^+ = 70.7$ (dense), $z^+\approx 17.7$}}
	\\ \vspace{-3.0\baselineskip}
	\subfloat[]{
		\includegraphics[trim=0 0 0 0,clip=true,scale=1.0]{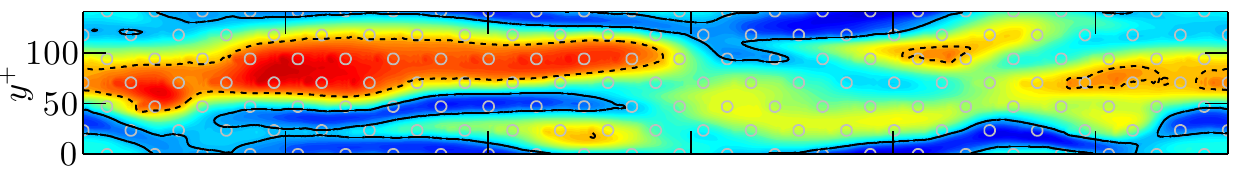}
	}
	\put(-4.6,1.12){\colorbox{white}{$\lambda^+ = 47.1$ (dense), $z^+\approx 18.4$}}
	\\ \vspace{-3.0\baselineskip}
	\subfloat[]{
		\includegraphics[trim=0 0 0 0,clip=true,scale=1.0]{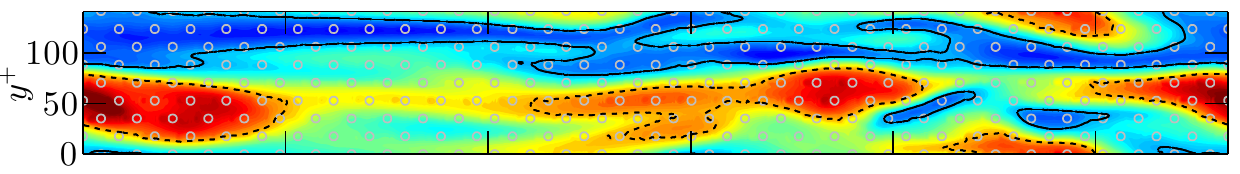}
	}
	\put(-4.6,1.12){\colorbox{white}{$\lambda^+ = 35.3$ (dense), $z^+\approx 18.8$}}
	\\ \vspace{-3.0\baselineskip}
	\subfloat[]{
		\includegraphics[trim=0 0 0 0,clip=true,scale=1.0]{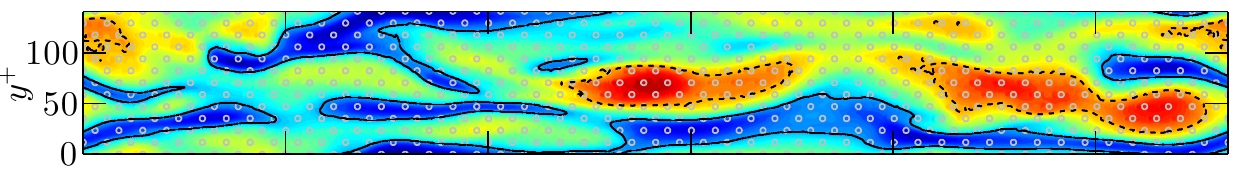}
	}
	\put(-4.6,1.12){\colorbox{white}{$\lambda^+ = 23.6$ (dense), $z^+\approx 19.8$}}
	\\ \vspace{-3.0\baselineskip}
	\subfloat[]{
		\includegraphics[trim=0 0 0 0,clip=true,scale=1.0]{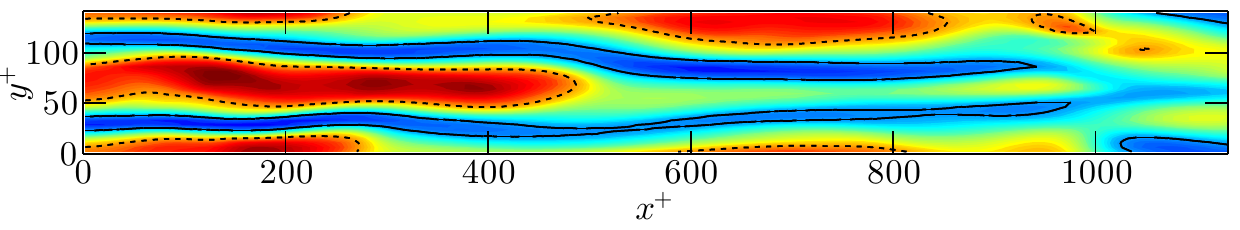}
	}
	\put(-5.8,1.7){\colorbox{white}{$\lambda^+ \rightarrow 0$ (dense limit), $z^+\approx 14.3+k^+$}}
	\\ \vspace{-2.5\baselineskip}
	\subfloat[]{
		\includegraphics[trim=0 0 0 0,clip=true,scale=1.0]{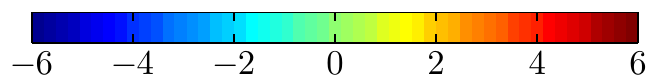}
	}
	\put(-3.6,-0.4){$u'^+$}
	\vspace{-0.75\baselineskip}
\caption{(Colour online) Contours of instantaneous streamwise turbulent fluctuations, $u'^+ = u(x,y,z,t)-\langle\overline{u}(x,y,z)\rangle$,  in a wall-parallel plane at the location of maximum $u'^+_{rms}$. Line contour threshold is set at $u'^+=\pm u'^+_{rms,max}$ (figure \ref{fig:FlucSol}\emph{a}).
Line styles are:
 solid line, negative fluctuations; 
dashed line, positive fluctuations.
Grey circles indicate the underlying roughness cross section at $z^+\approx7.5$ (2.5 wall-units below roughness crest).
}
	\label{fig:UzContourStreaks}
\end{figure}

Snapshots of the streamwise turbulent fluctuations $u'^+$ are shown in figure \ref{fig:UzContourStreaks}, where the turbulent fluctuation is the deviation from the temporally and spatially averaged mean, $u'(x,y,z,t) = \langle\overline{u}\rangle(z) - u(x,y,z,t)$.
The long streaky structure over the smooth wall is  still clearly present in the roughness cases, albeit at a higher wall-normal position.
Streaky structures over rough walls have been investigated by multiple authors \citep{Raupach91,Lee11,DeMarchis15}, and it should be noted that it is likely that they are observed over the current surface as it is a small ($k^+=10$) roughness in the transitionally rough regime.
While difficult to draw conclusions from a single instantaneous snapshot, the streaky structure over the sparse regime ($\lambda^+=113$) and the first dense regime case ($\lambda^+=70.7$) appears to be somewhat disrupted compared to the smooth wall. The structures in the sparse regime do not appear to extend as far in the streamwise direction, suggesting that the peak streamwise energy is located at shorter wavelengths, although a longer streamwise channel domain would be required to assess this claim further. The peak streamwise turbulence intensity (figure \ref{fig:FlucSol}\emph{a}) indicates that the intensity of the fluctuations are reduced for these sparser cases, although it is challenging to see this effect in one snapshot. However, the denser roughness cases ($\lambda^+\le47.1$) show a more similar structure to the smooth wall, in that the high- and low-speed velocity streaks becoming increasingly elongated. In the limit of $\lambda^+\rightarrow0$, the smooth-wall streaky structure is clearly re-established, although now $k^+$ wall-units further from the origin.
It is important to emphasise that these qualitative descriptions of a single snapshot are made to give a physical visualisation to the changes in flow; the evidence comes from the previous figures, particularly figures \ref{fig:velFluc}(\emph{b}) and \ref{fig:EuuContour}.

The picture that emerges is one where the near-wall cycle is increasingly weakened by the roughness with increasing solidity, while in the sparse regime. At the crossover between sparse and dense regimes, the near-wall cycle is most damped. As solidity increases beyond this value ($\Lambda\gtrsim0.18$), into the dense regime, the near-wall cycle continues to be pushed outwards, but re-strengthens as the new smooth limit at $\Lambda\rightarrow\infty$ is approached. Commensurate with this observation, the minimum viscous drag (and maximum pressure drag) is attained at the crossover between sparse and dense regimes.

The conclusions made above regarding the near-wall cycle would only apply to the transitionally rough regime. In the fully rough regime, the near-wall cycle is believed to be completely destroyed and the inertial logarithmic region starts at the roughness crest. The cycle would, however, still exist in the limits of $\lambda\rightarrow0$ and $\lambda\rightarrow\infty$ as these are effectively smooth walls. Additional numerical simulations of the fully rough regime would be required to understand what mechanism causes the reduction in $\Delta U^+$ in the fully rough dense regime. The mean momentum balance analysis could still be applied however the interpretation of the $T_1$ and $T_2$ terms would change in this regime.


\section{Conclusions}
\label{sect:discConc}

The sparse and dense regimes of roughness were investigated using direct numerical simulations of flow over three-dimensional transitionally rough sinusoidal roughness. The minimal-span channel technique, recently investigated by \cite{Chung15} for rough-wall flows, was used. The minimal-span channel was seen to accurately predict the roughness function $\Delta U^+$ for both sparse and dense roughness when compared to full-span channels. Moreover, analysis of second-order turbulence statistics showed that the root-mean-square streamwise and wall-normal velocity fluctuations were accurately captured by the minimal-span channel, especially within the roughness cavities.

The dense regime of roughness was found to occur when the solidity was greater than approximately 0.15--0.18. In this regime, the velocity fluctuations within the roughness cavities decreased, although were not negligible even for the densest case simulated ($\Lambda=0.54$). An analysis of the mean momentum balance enabled the roughness function to be decomposed into two contributions; $T_1$, the difference in streamwise velocity at the roughness crest between smooth-wall and rough-wall flows, and $T_2$, the integrated difference in Reynolds shear stress between the two flows.
This revealed that the primary reason for the reduction in the roughness function that is seen in the dense regime is due to the reduction in Reynolds shear stress above the roughness elements, i.e. due to $T_2$. 

The limit of $\Lambda\rightarrow\infty$ was assumed to have negligible flow within the roughness cavities, so was modelled as a smooth wall in which the wall was located at the roughness crest.
A Stokes flow simulation over a dense cuboid rough surface with $\Lambda=4.44$ was performed and showed near-zero velocity at the roughness crest. As such, the data showed good agreement with the classical parabolic velocity profile where the wall was offset to the roughness crests, supporting this modelling assumption.
The  near-wall turbulent statistics of the rough-wall flows also appear to be tending towards this offset smooth-wall limit.
A simple extrapolation of the log-linear $T_1$ contribution indicated that the asymptotic limit of $\Lambda\rightarrow\infty$ is reached when $\Lambda\approx4.2$, corresponding to $\lambda^+\approx3$. This supports the view that roughness effects are not felt when the roughness wavelength $\lambda\rightarrow\nu/U_\tau$.

As solidity increases, it appears that the near-wall cycle is being pushed up away from the roughness. From the peak streamwise turbulence intensity (figure \ref{fig:FlucSol}) and energy spectra (figure \ref{fig:EuuContour}), the energy peak is moving away from the wall for both sparse and dense roughness. Increasing solidity in the sparse regime leads to a reduction in the peak energy, while increasing solidity in the dense regime causes an increase in peak energy (figure \ref{fig:FlucSol}\emph{a}). Compared to the smooth wall, the dense roughness cases have less turbulent energy in the region immediately above the crest of the roughness ($k^+\le z^+\lesssim17$, figure \ref{fig:EuuContour}\emph{b}--\emph{d}) which was previously occupied by the near-wall cycle. The absence of streamwise energy in this region corresponds to a reduced Reynolds shear stress, and the $T_2$ term is therefore negative. From the analysis of the mean momentum balance, this then causes the reduction in $\Delta U^+$ for the dense regime, $\Lambda> 0.15$.


\section*{Acknowledgements}
The authors would like to gratefully acknowledge the financial support of the Australian Research Council and the Bushfire and Natural Hazards Cooperative Research Council. Computational time was granted under the Victoria Life Sciences Computational Initiative (VLSCI), which is supported by the Victorian Government, Australia.


\section*{Appendix. Mean momentum balance for different $Re_\tau$}\label{appA}
The analysis of the mean momentum balanced presented in \S\ref{sect:mmb} assumed the smooth- and rough-wall flows have matched friction Reynolds numbers, $Re_\tau$. The simulations performed in this study do not necessarily meet this requirement, with the variation in $Re_\tau$ being up to $\pm2\%$ (table \ref{tab:sims}). We will therefore rigorously perform the analysis without the matched $Re_\tau$ assumption. Starting from the smooth- and rough-wall mean momentum balance equations, we have
\begin{subequations}
\label{eqn:mmb_app}
\begin{align}
\tau_{uwS} + \nu_S \df{U_S}{ z_S} &= U_{\tau S}^2 \frac{h_S-z_S}{h_S}  \\
\tau_{uwR} + \nu_R \df{U_R}{ z_R} &= U_{\tau R}^2 \frac{h_R-z_R}{h_R},
\end{align}
\end{subequations}
where subscripts $S$ and $R$ denote smooth- and rough-wall flows, respectively. Superscript $+$ implies non-dimensionalisation on the smooth wall viscosity, $\nu_S$, and friction velocity, $U_{\tau S}$. Superscript $\star$ implies non-dimensionalisation on the rough wall viscosity $\nu_R$, and friction velocity $U_{\tau R}$. Firstly we will assume matched inner-normalised position, $z_S^+=z_R^\star$, which implies
\begin{eqnarray}
\frac{z_S U_{\tau S}}{\nu_S} = \frac{z_R U_{\tau R}}{\nu_R}   \Rightarrow \frac{z_S}{h_S} = \frac{h_R^\star}{h_S^+}\frac{z_R}{h_R}. \nonumber
\end{eqnarray}
Let $\alpha = h_R^\star/h_S^+$, the difference in friction Reynolds numbers. Put $\alpha=1+\varepsilon$, where $\varepsilon = h_R^\star/h_S^+-1=(h_R^\star-h_S^+)/h_S^+$. The mean-momentum balance (\ref{eqn:mmb_app}) can then be equated to obtain
\begin{eqnarray}
\label{eqn:equatedMMB}
\df{U_S^+}{ z_S^+} -\alpha\df{U_R^\star}{ z_R^\star} &=&
\alpha\tau_{uwR}^\star- \tau_{uwS}^++1 -\alpha .
\end{eqnarray}

We will now integrate from the peak of the roughness $k^\star$ to the rough-wall channel centre $h_R^\star$ and rearrange for the roughness function, $\Delta U^+$.
\begin{subequations}
\label{eqn:app_final}
\begin{align}
\Delta U^+ &= U_S^+(z_R^\star=h_R^\star) - U_R^\star(z_R^\star=h_R^\star)\nonumber\\
&= U_S^+(z_R^\star=k^\star) -  U_R^\star(z_R^\star=k^\star)\label{eqn:T1_app} \\
&+ \int_{z_R^\star=k^\star}^{z_R^\star=h_R^\star}\left(\tau_{uwR}^\star(z_R^\star) - \tau_{uwS}^+(z_R^\star)\right)\dd z_R^\star \label{eqn:T2_app}\\
&+\epsilon\left(U_R^\star(z_R^\star=h_R^\star)-U_R^\star(z_R^\star=k^\star)+\int_{z_R^\star=k^\star}^{z_R^\star=h_R^\star}\tau_{uwR}^\star\id z_R^\star-(h_R^\star-k^\star) \right)\nonumber
\end{align}
\end{subequations}

This is similar to (\ref{eqn:DU_decomp}) derived in \S\ref{sect:mmb}, where $T_1$ corresponds to (\ref{eqn:T1_app})  and $T_2$ corresponds to (\ref{eqn:T2_app}),  however the non-dimensionalisation uses the respective  smooth- and rough-wall friction velocity and viscosity.
 There is also an additional error term that is proportional to the difference in friction Reynolds numbers, $\epsilon = h_R^\star/h_S^+-1$. For the current data, $|\varepsilon|\lesssim 0.02$ and so this term is negligible.

\setlength{\unitlength}{1cm}
\begin{figure}
\centering
 \captionsetup[subfigure]{labelformat=empty}

	\subfloat[]{
		\includegraphics[width=0.49\textwidth]{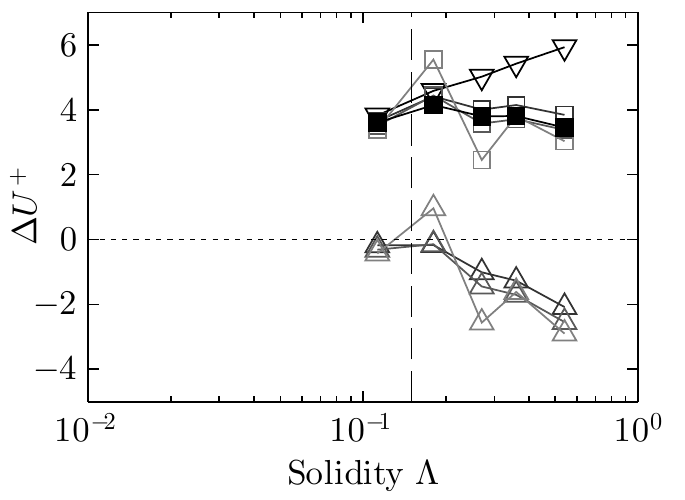}
	}
	\put(-1.05,4.25){$T_1$}
	\put(-1.05,1.5){$T_2$}
	\vspace{-2.0\baselineskip}
\caption{Roughness function decomposition (\ref{eqn:app_final}), for different wall-normal integration positions $z_i^\star$  in minimal-span channel flow. 
Line styles:
\protect\raisebox{0.4ex}{\protect\scalebox{1.0}{$\bigtriangledown$}}, $T_1$ (\ref{eqn:T1_app});
$\bigtriangleup$, $T_2$ (\ref{eqn:T2_int});
$\square$, $T_1 + T_2$;
$\blacksquare$, $\Delta U^+$ computed from mean velocity profile.
Black to grey lines refer to $z_i^\star = (25,50,180)$.
Vertical dashed line at $\Lambda = 0.15$ shows start of dense regime.
}
\label{fig:T2_int}
\end{figure}

The remaining issue is that the Reynolds shear stress should collapse in the outer layer for matched friction Reynolds numbers, according to Townsend's outer-layer similarity hypothesis. This means that the difference between  $\tau_{uwS}^+$ and $\tau_{uwR}^\star$ in (\ref{eqn:T2_app}) should be zero for $z^\star\gg 1$.
 However if $|\varepsilon|>0$ this is not necessarily the case. If we consider some wall-normal position $z_i$, then the integral of (\ref{eqn:T2_app}) can be split as,
\begin{eqnarray}
\label{eqn:T2_int}
\int_{z_R^\star=k^\star}^{z_R^\star=z_i^\star}\left(\tau_{uwR}^\star(z_R^\star) - \tau_{uwS}^+(z_R^\star)\right)\dd z_R^\star
+  \int_{z_R^\star=z_i^\star}^{z_R^\star=h_R^\star}\left(\tau_{uwR}^\star(z_R^\star) - \tau_{uwS}^+(z_R^\star)\right)\dd z_R^\star.
\end{eqnarray}
Now if $z_i$ is reasonably far away from the wall, we assume that the second term should be negligible.  The question is then what value of $z_i^\star$ is appropriate.

Different values for $z_i^\star$ are used in calculating (\ref{eqn:T2_int}) with the second term ignored, and are shown in figure \ref{fig:T2_int}. It can be seen that using $z_i^\star = 25$ or $z_i^\star=50$ results in a reasonable approximation of $\Delta U^+$. However, as explained, the outer-layer (second) term of (\ref{eqn:T2_int}) can dominate for $z_i^\star=h_R^\star$, leading to a large error in $T_1 + T_2$. It is important to emphasise that all values of $z_i^\star$ still show that the dense roughness regime has a reduction in the roughness function primarily due to the reduction in Reynolds shear stress. The conclusions made in this paper are therefore independent of $z_i^\star$. The choice of $z_i^\star$ is selected as $z_i^\star=50$.

\bibliographystyle{jfm}
\bibliography{bibliography}

\begin{thebibliography}{48}
\expandafter\ifx\csname natexlab\endcsname\relax\def\natexlab#1{#1}\fi

\bibitem[Antonia \& Krogstad(2001)]{Antonia01}
{\sc Antonia, R.~A. \& Krogstad, P.-{\AA}.} 2001 Turbulence structure in
  boundary layers over different types of surface roughness. {\em Fluid Dyn.
  Res.\/} {\bf 28}~(2), 139--157.

\bibitem[Beavers \& Joseph(1967)]{Beavers67}
{\sc Beavers, G.~S. \& Joseph, D.~D.} 1967 Boundary conditions at a naturally
  permeable wall. {\em J. Fluid Mech.\/} {\bf 30}~(01), 197--207.

\bibitem[Breugem {\em et~al.\/}(2006)Breugem, Boersma \&
  Uittenbogaard]{Breugem06}
{\sc Breugem, W.~P., Boersma, B.~J. \& Uittenbogaard, R.~E.} 2006 The influence
  of wall permeability on turbulent channel flow. {\em J. Fluid Mech.\/} {\bf
  562}~(1), 35--72.

\bibitem[Chan {\em et~al.\/}(2015)Chan, MacDonald, Chung, Hutchins \&
  Ooi]{Chan15}
{\sc Chan, L., MacDonald, M., Chung, D., Hutchins, N. \& Ooi, A.} 2015 A
  systematic investigation of roughness height and wavelength in turbulent pipe
  flow in the transitionally rough regime. {\em J. Fluid Mech.\/} {\bf 771},
  743--777.

\bibitem[Chung {\em et~al.\/}(2015)Chung, Chan, MacDonald, Hutchins \&
  Ooi]{Chung15}
{\sc Chung, D., Chan, L., MacDonald, M., Hutchins, N. \& Ooi, A.} 2015 A fast
  direct numerical simulation method for characterising hydraulic roughness.
  {\em J. Fluid Mech.\/} {\bf 773}, 418--431.

\bibitem[Coceal {\em et~al.\/}(2006)Coceal, Thomas, Castro \&
  Belcher]{Coceal06}
{\sc Coceal, O., Thomas, T.~G., Castro, I.~P. \& Belcher, S.~E.} 2006 Mean flow
  and turbulence statistics over groups of urban-like cubical obstacles. {\em
  Boundary-Layer Meteorol.\/} {\bf 121}~(3), 491--519.

\bibitem[De~Marchis {\em et~al.\/}(2015)De~Marchis, Milici \&
  Napoli]{DeMarchis15}
{\sc De~Marchis, M., Milici, B. \& Napoli, E.} 2015 Numerical observations of
  turbulence structure modification in channel flow over {2D} and {3D} rough
  walls. {\em Int. J. Heat Fluid Fl.\/} {\bf 56}, 108--123.

\bibitem[Efros \& Krogstad(2011)]{Efros11}
{\sc Efros, V. \& Krogstad, P.-{\AA}.} 2011 Development of a turbulent boundary
  layer after a step from smooth to rough surface. {\em Exp. Fluids\/} {\bf
  51}~(6), 1563--1575.

\bibitem[Flack \& Schultz(2010)]{Flack10}
{\sc Flack, K.~A. \& Schultz, M.~P.} 2010 Review of hydraulic roughness scales
  in the fully rough regime. {\em J. Fluids Eng.\/} {\bf 132}~(4), 041203.

\bibitem[Flack \& Schultz(2014)]{Flack14}
{\sc Flack, K.~A. \& Schultz, M.~P.} 2014 Roughness effects on wall-bounded
  turbulent flows. {\em Phys. Fluids\/} {\bf 26}~(10), 101305.

\bibitem[Flack {\em et~al.\/}(2012)Flack, Schultz \& Rose]{Flack12}
{\sc Flack, K.~A., Schultz, M.~P. \& Rose, W.~B.} 2012 The onset of roughness
  effects in the transitionally rough regime. {\em Int. J. Heat Fluid Fl.\/}
  {\bf 35}, 160--167.

\bibitem[Flores \& Jim{\'e}nez(2010)]{Flores10}
{\sc Flores, O. \& Jim{\'e}nez, J.} 2010 Hierarchy of minimal flow units in the
  logarithmic layer. {\em Phys. Fluids\/} {\bf 22}~(7), 071704.

\bibitem[Fukagata {\em et~al.\/}(2002)Fukagata, Iwamoto \& Kasagi]{Fukagata02}
{\sc Fukagata, K., Iwamoto, K. \& Kasagi, N.} 2002 Contribution of {R}eynolds
  stress distribution to the skin friction in wall-bounded flows. {\em Phys.
  Fluids\/} {\bf 14}~(11), L73--L76.

\bibitem[Garc{\'\i}a-Mayoral \& Jim{\'e}nez(2011)]{Mayoral11jfm}
{\sc Garc{\'\i}a-Mayoral, R. \& Jim{\'e}nez, J.} 2011 Hydrodynamic stability
  and breakdown of the viscous regime over riblets. {\em J. Fluid Mech.\/} {\bf
  678}, 317--347.

\bibitem[Grimmond \& Oke(1999)]{Grimmond99}
{\sc Grimmond, C. S.~B. \& Oke, T.~R.} 1999 Aerodynamic properties of urban
  areas derived from analysis of surface form. {\em J. Appl. Meteorol.\/} {\bf
  38}~(9), 1262--1292.

\bibitem[Hagishima {\em et~al.\/}(2009)Hagishima, Tanimoto, Nagayama \&
  Meno]{Hagishima09}
{\sc Hagishima, A., Tanimoto, J., Nagayama, K. \& Meno, S.} 2009 Aerodynamic
  parameters of regular arrays of rectangular blocks with various geometries.
  {\em Boundary-Layer Meteorol.\/} {\bf 132}~(2), 315--337.

\bibitem[Ham \& Iaccarino(2004)]{Ham04}
{\sc Ham, F. \& Iaccarino, G.} 2004 Energy conservation in collocated
  discretization schemes on unstructured meshes. In {\em Annual Research Briefs
  2004\/}, pp. 3--14. Center for Turbulence Research, Stanford University/NASA
  Ames.

\bibitem[Hamilton {\em et~al.\/}(1995)Hamilton, Kim \& Waleffe]{Hamilton95}
{\sc Hamilton, J.~M., Kim, J. \& Waleffe, F.} 1995 Regeneration mechanisms of
  near-wall turbulence structures. {\em J. Fluid Mech.\/} {\bf 287}, 317--348.

\bibitem[Hwang(2013)]{Hwang13}
{\sc Hwang, Y.} 2013 Near-wall turbulent fluctuations in the absence of wide
  outer motions. {\em J. Fluid Mech.\/} {\bf 723}, 264--288.

\bibitem[Jim{\'e}nez(2004)]{Jimenez04}
{\sc Jim{\'e}nez, J.} 2004 Turbulent flows over rough walls. {\em Annu. Rev.
  Fluid Mech.\/} {\bf 36}, 173--196.

\bibitem[Jim{\'e}nez \& Moin(1991)]{Jimenez91}
{\sc Jim{\'e}nez, J. \& Moin, P.} 1991 The minimal flow unit in near-wall
  turbulence. {\em J. Fluid Mech.\/} {\bf 225}, 213--240.

\bibitem[Jim{\'e}nez \& Pinelli(1999)]{Jimenez99}
{\sc Jim{\'e}nez, J. \& Pinelli, A.} 1999 The autonomous cycle of near-wall
  turbulence. {\em J. Fluid Mech.\/} {\bf 389}, 335--359.

\bibitem[Kanda {\em et~al.\/}(2004)Kanda, Moriwaki \& Kasamatsu]{Kanda04}
{\sc Kanda, M., Moriwaki, R. \& Kasamatsu, F.} 2004 Large-eddy simulation of
  turbulent organized structures within and above explicitly resolved cube
  arrays. {\em Boundary-Layer Meteorol.\/} {\bf 112}~(2), 343--368.

\bibitem[Krumbein \& Monk(1943)]{Krumbein43}
{\sc Krumbein, W.~C. \& Monk, G.~D.} 1943 Permeability as a function of the
  size parameters of unconsolidated sand. {\em Trans. Inst. Min. Met. Engrs.\/}
  {\bf 151}~(01), 153--163.

\bibitem[Lee {\em et~al.\/}(2011)Lee, Sung \& Krogstad]{Lee11}
{\sc Lee, J.~H., Sung, H.~J. \& Krogstad, P.-{\AA}.} 2011 Direct numerical
  simulation of the turbulent boundary layer over a cube-roughened wall. {\em
  J. Fluid Mech.\/} {\bf 669}, 397--431.

\bibitem[Leonardi \& Castro(2010)]{Leonardi10}
{\sc Leonardi, S. \& Castro, I.~P.} 2010 Channel flow over large cube
  roughness: a direct numerical simulation study. {\em J. Fluid Mech.\/} {\bf
  651}, 519--539.

\bibitem[Leonardi {\em et~al.\/}(2007)Leonardi, Orlandi \& Antonia]{Leonardi07}
{\sc Leonardi, S., Orlandi, P. \& Antonia, R.~A.} 2007 Properties of d-and
  k-type roughness in a turbulent channel flow. {\em Phys. Fluids\/} {\bf 19},
  125101.

\bibitem[Lien {\em et~al.\/}(2005)Lien, Yee \& Wilson]{Lien05}
{\sc Lien, F.~S., Yee, E. \& Wilson, J.~D.} 2005 Numerical modelling of the
  turbulent flow developing within and over a 3-{D} building array, part {II}:
  a mathematical foundation for a distributed drag force approach. {\em
  Boundary-Layer Meteorol.\/} {\bf 114}~(2), 245--285.

\bibitem[Lozano-Dur{\'a}n \& Jim{\'e}nez(2014)]{Lozano14}
{\sc Lozano-Dur{\'a}n, A. \& Jim{\'e}nez, J.} 2014 Effect of the computational
  domain on direct simulations of turbulent channels up to ${R}e_\tau$= 4200.
  {\em Phys. Fluids\/} {\bf 26}~(1), 011702.

\bibitem[Macdonald(2000)]{Macdonald00}
{\sc Macdonald, R.~W.} 2000 Modelling the mean velocity profile in the urban
  canopy layer. {\em Boundary-Layer Meteorol.\/} {\bf 97}~(1), 25--45.

\bibitem[Macdonald {\em et~al.\/}(1998)Macdonald, Griffiths \&
  Hall]{Macdonald98}
{\sc Macdonald, R.~W., Griffiths, R.~F. \& Hall, D.~J.} 1998 An improved method
  for the estimation of surface roughness of obstacle arrays. {\em Atmos.
  Environ.\/} {\bf 32}~(11), 1857--1864.

\bibitem[Mahesh {\em et~al.\/}(2004)Mahesh, Constantinescu \& Moin]{Mahesh04}
{\sc Mahesh, K., Constantinescu, G. \& Moin, P.} 2004 A numerical method for
  large-eddy simulation in complex geometries. {\em J. Comput. Phys.\/} {\bf
  197}, 215–--240.

\bibitem[Millward-Hopkins {\em et~al.\/}(2011)Millward-Hopkins, Tomlin, Ma,
  Ingham \& Pourkashanian]{MillwardHopkins11}
{\sc Millward-Hopkins, J.~T., Tomlin, A.~S., Ma, L., Ingham, D. \&
  Pourkashanian, M.} 2011 Estimating aerodynamic parameters of urban-like
  surfaces with heterogeneous building heights. {\em Boundary-Layer
  Meteorol.\/} {\bf 141}~(3), 443--465.

\bibitem[Moin \& Kim(1982)]{Moin82}
{\sc Moin, P. \& Kim, J.} 1982 Numerical investigation of turbulent channel
  flow. {\em J. Fluid Mech.\/} {\bf 118}, 341--377.

\bibitem[Napoli {\em et~al.\/}(2008)Napoli, Armenio \& De~Marchis]{Napoli08}
{\sc Napoli, E., Armenio, V. \& De~Marchis, M.} 2008 The effect of the slope of
  irregularly distributed roughness elements on turbulent wall-bounded flows.
  {\em J. Fluid Mech.\/} {\bf 613}, 385--394.

\bibitem[Nikuradse(1933)]{Nikuradse33}
{\sc Nikuradse, J.} 1933 Laws of flow in rough pipes. {T}ranslation from
  {G}erman published 1950 as {NACA} {T}ech. {M}emo. 1292.

\bibitem[Oke(1988)]{Oke88}
{\sc Oke, T.~R} 1988 Street design and urban canopy layer climate. {\em Energ.
  Buildings\/} {\bf 11}~(1), 103--113.

\bibitem[Orlandi {\em et~al.\/}(2006)Orlandi, Leonardi \&
  Antonia]{Orlandi06jfm}
{\sc Orlandi, P., Leonardi, S. \& Antonia, R.~A.} 2006 Turbulent channel flow
  with either transverse or longitudinal roughness elements on one wall. {\em
  J. Fluid Mech.\/} {\bf 561}, 279--305.

\bibitem[Placidi \& Ganapathisubramani(2015)]{Placidi15}
{\sc Placidi, M. \& Ganapathisubramani, B.} 2015 Effects of frontal and plan
  solidities on aerodynamic parameters and the roughness sublayer in turbulent
  boundary layers. {\em J. Fluid Mech.\/} {\bf 782}, 541--566.

\bibitem[Raupach(1994)]{Raupach94}
{\sc Raupach, M.~R.} 1994 Simplified expressions for vegetation roughness
  length and zero-plane displacement as functions of canopy height and area
  index. {\em Boundary-Layer Meteorol.\/} {\bf 71}~(1-2), 211--216.

\bibitem[Raupach {\em et~al.\/}(1991)Raupach, Antonia \&
  Rajagopalan]{Raupach91}
{\sc Raupach, M.~R., Antonia, R.~A. \& Rajagopalan, S.} 1991 Rough-wall
  turbulent boundary layers. {\em Appl. Mech. Rev.\/} {\bf 44}, 1--25.

\bibitem[Rosti {\em et~al.\/}(2015)Rosti, Cortelezzi \& Quadrio]{Rosti15}
{\sc Rosti, M.~E., Cortelezzi, L. \& Quadrio, M.} 2015 Direct numerical
  simulations of turbulent channel flow over porous walls. {\em J. Fluid
  Mech.\/} {\bf 784}, 396--442.

\bibitem[Schlichting(1936)]{Schlichting36}
{\sc Schlichting, H.} 1936 Experimental investigation of the problem of surface
  roughness. {\em Ing.-Arch.\/} {\bf 7}, 1--34, {T}ranslation from {G}erman
  published 1937 as {NACA} {T}ech. {M}emo. 823.

\bibitem[Schultz \& Flack(2009)]{Schultz09}
{\sc Schultz, M.~P. \& Flack, K.~A.} 2009 Turbulent boundary layers on a
  systematically varied rough wall. {\em Phys. Fluids\/} {\bf 21}, 015104.

\bibitem[Spalart \& McLean(2011)]{Spalart11}
{\sc Spalart, P.~R. \& McLean, J.~D.} 2011 Drag reduction: enticing turbulence,
  and then an industry. {\em Phil. Trans. R. Soc. A\/} {\bf 369}~(1940),
  1556--1569.

\bibitem[Townsend(1976)]{Townsend76}
{\sc Townsend, A.~A.} 1976 {\em The {S}tructure of {T}urbulent {S}hear
  {F}low\/}, 2nd edn. Cambridge University Press.

\bibitem[Waigh \& Kind(1998)]{Waigh98}
{\sc Waigh, D.~R. \& Kind, R.~J.} 1998 Improved aerodynamic characterization of
  regular three-dimensional roughness. {\em AIAA J.\/} {\bf 36}~(6),
  1117--1119.

\bibitem[Yang {\em et~al.\/}(2016)Yang, Sadique, Mittal \& Meneveau]{Yang16}
{\sc Yang, X. I.~A., Sadique, J., Mittal, R. \& Meneveau, C.} 2016 Exponential
  roughness layer and analytical model for turbulent boundary layer flow over
  rectangular-prism roughness elements. {\em J. Fluid Mech.\/} {\bf 789},
  127--165.

\end{thebibliography}

\end{document}